\documentclass[
reprint,
superscriptaddress,
amsmath,
amssymb,
aps,
prxquantum,
longbibliography,
nofootinbib
]{revtex4-2}

\usepackage{graphicx}
\usepackage{dcolumn}
\usepackage{bm}
\usepackage{xcolor}
\usepackage{hyperref}
\hypersetup{
    colorlinks=true,
    linkcolor=blue,
    filecolor=magenta,
    urlcolor=cyan,
    citecolor=blue,
}

\newcommand{\avg}[1]{\left\langle #1 \right\rangle}
\newcommand{\var}[1]{\mathrm{Var}\!\left( #1 \right)}
\newcommand{\snr}{\mathrm{SNR}}

\begin{document}

\title{Sub-Vacuum Jamming for Secure Communication}

\author{Sh.~Barzanjeh}
\email{shabir.barzanjeh@ucalgary.ca}
\affiliation{Institute for Quantum Science and Technology and Department of Physics and Astronomy, University of Calgary, 2500 University Drive NW, Calgary, AB T2N 1N4, Canada}
\author{A.~Poostindouz}
\affiliation{Department of Computer Science, University of Calgary, 2500 University Drive NW, Calgary, AB T2N 1N4, Canada}
\author{B.~C.~Sanders}
\affiliation{Institute for Quantum Science and Technology and Department of Physics and Astronomy, University of Calgary, 2500 University Drive NW, Calgary, AB T2N 1N4, Canada}

\date{\today}

\begin{abstract}
Artificial-noise jamming improves physical-layer security by degrading an eavesdropper's channel, but the injected noise also interferes with the legitimate receiver. We introduce a correlated-jamming protocol based on classical and quantum correlations that suppresses this self-interference while preserving the jamming penalty experienced by the eavesdropper. Bob broadcasts one mode of a correlated two-mode source and retains the second as a local reference. Eve, who has no access to the retained mode, receives the full thermal jamming field, whereas Bob uses an optimized joint measurement to cancel its correlated fluctuations.
The residual self-jamming noise is governed by the conditional variance of the transmitted jamming quadrature given the retained reference. Any classically correlated source is bounded by the vacuum-noise floor and therefore restores Bob, at best, to his unjammed receiver noise. An entangled two-mode squeezed source surpasses this limit and produces sub-vacuum residual noise, with the maximum quantum advantage set by the transmissivity of the jamming path back to Bob. In the bright-source regime, this advantage becomes a fixed reduction in Bob's noise floor, independent of the signal and jamming powers.
Within a bounded-collection wiretap model, correlated jamming preserves positive secrecy even when Eve has a stronger direct channel than Bob and remains effective against collective measurements on Eve's Gaussian output states. The protocol supports bright classical message transmission and requires no end-to-end quantum channel. Possible applications include shot-noise-limited optical communication, cryogenic microwave networks, and covert or power-constrained secure links.

\end{abstract}

\maketitle

\section{Introduction}\label{sec:intro}

Any field launched into a lossy medium can be tapped by anyone within its
light cone. Optical fibers, free-space links and wireless broadcasts are all
vulnerable in this way, and the resulting security problem is modeled by the
wiretap channel \cite{Wyner1975,Csiszar1978}. A sender (Alice) communicates
with a legitimate receiver (Bob) while an eavesdropper (Eve) observes a noisy
version of the same transmission. Communication remains possible at a rate,
the secrecy capacity, at which Eve's leakage rate vanishes. This guarantee assumes
nothing about Eve's computational power, which is what makes physical-layer
security a complement to application-layer cryptography rather than a
replacement for it. Cryptographic security is computational, and therefore
contingent on certain mathematical problems remaining hard, an assumption
under growing pressure from quantum algorithms \cite{NielsenChuang2010}. For
the Gaussian channels of optical and radio-frequency links, the secrecy
capacity is the difference between the Alice--Bob and Alice--Eve capacities,
and it is positive precisely when Bob's channel is the better one
\cite{LeungYanCheong1978}.

This condition represents a central limitation of the entire approach because it makes secrecy dependent on geometry. An eavesdropper located closer to the transmitter than the intended receiver, or equipped with a more sensitive antenna, can reduce the secrecy capacity to zero without the legitimate parties necessarily detecting the loss of security. A common countermeasure is cooperative jamming, also known as artificial-noise injection \cite{GoelNegi2008,Tekin2008,Dong2009,Cumanan2017,Xu2019,Jameel2019,Mucchi2021}. In this approach, a helper node, often the receiver itself, injects noise into the channel to create an effective channel advantage for the legitimate link.

The main challenge is that the artificial noise affects the legitimate receiver as well as the eavesdropper. Classical strategies address this problem by steering the noise into the null space of the legitimate channel using antenna arrays,  exploiting favorable relay geometries \cite{Xu2019,Jameel2019,Cumanan2017}, or suppressing the jammer's own emission through self-interference cancellation, as developed for in-band full-duplex radio \cite{Zheng2013,Sabharwal2014}. These methods require accurate channel-state information, additional hardware, or both, and their performance deteriorates rapidly under imperfect channel estimation or incomplete interference suppression.

The limitation becomes more fundamental in the optical domain. An artificial-noise field contains both classical excess fluctuations and irreducible quantum shot noise. A receiver with access to a classical record of the injected noise can subtract the classical component, but not the associated quantum fluctuations. Classical resources therefore impose a fundamental limit on how effectively a jammer can remove its own interference.

Quantum key distribution (QKD) achieves information-theoretic secrecy without
computational assumptions
\cite{BB84,Ekert1991,Gisin2002,Scarani2009,Pirandola2020} , and has been carried into field-deployed networks that combine metropolitan fiber with satellite-to-ground links \cite{Chen2021Network}. Its security relies 
on the laws of quantum mechanics while the adversary is granted full control of
the communication channel. Continuous-variable QKD encodes the key in the
quadratures of squeezed or coherent optical fields and reads it out by
homodyne detection
\cite{Ralph1999,Hillery2000,Reid2000,Cerf2001,GottesmanPreskill2001,Grosshans2002,Silberhorn2002,Grosshans2003,Weedbrook2004,Weedbrook2012},
which places it technologically close to coherent optical communication and has
supported demonstrations over hundreds of kilometers of fiber with composable
security \cite{Zhang2020,Jain2022,Hajomer2024}, as well as chip-scale transmitters and
receivers built in silicon photonics \cite{ZhangChip2019}.

This adversarial model comes with operational constraints. The transmitted
states must remain weak and fragile and must travel on dedicated low-noise
channels; secret-key rates fall with distance and are ultimately capped by the
repeaterless rate--loss limit \cite{Pirandola2017, Wang2022TF}; and coexistence with
bright classical traffic remains difficult \cite{Diamanti2016}. Between
conventional cryptography and fully quantum communication lies a third
approach, in which quantum light does not carry the message but serves instead
as a local resource that strengthens the security of an ordinary classical
signal. Related ideas appear in keyed communication using quantum noise
\cite{Barbosa2003}, quantum-illumination-protected communication
\cite{Shapiro2009}, floodlight QKD \cite{Zhuang2016}, quantum data locking
\cite{Lupo2014}, and quantum-secure covert communication \cite{Bash2015}.

This paper develops such a hybrid approach based on correlated jamming. Alice
transmits an ordinary bright coherent signal, so no quantum state carrying the
message ever propagates through the channel. Bob generates either a
classically correlated source or a two-mode squeezed-vacuum (TMSV) source
\cite{Weedbrook2012}, broadcasts one mode as the jamming field,
and retains the correlated idler.

Taken on its own, the jamming field has the same thermal statistics for both
sources. Eve therefore receives Alice's signal on top of a broadband noise
background whose variance grows with the jamming power, and, having no system
correlated with that background, she cannot estimate and subtract its
fluctuations. Spectral filtering does not help her either, because the jamming
field is matched to the signal band. The two sources thus impose the same
penalty at Eve's receiver.

Bob is in a different position because he holds the idler. Interfering it with
the received field---or, equivalently, combining the two homodyne
photocurrents with optimized weights---cancels the part of the jamming noise
that is correlated with the idler. A symmetric classical source leaves a
residual $3$~dB above the shot-noise level. An idler-optimized classical
source approaches the unjammed vacuum floor in the ideal limit but cannot fall
below it, which is what makes it the appropriate classical benchmark. A TMSV
source passes that limit and suppresses the residual noise below the vacuum
level. As the jamming power grows, Eve's noise penalty grows with it, while
Bob's residual noise settles at a level fixed by the source correlations and
by the loss along the jamming path. Unlike the optimal classical strategy, the
quantum protocol never requires Bob to keep an accurate classical record of
the waveform he transmitted.

The quantum advantage does not require entanglement to survive propagation. As
in quantum illumination
\cite{Lloyd2008,Tan2008,Barzanjeh2015,Barzanjeh2020,Shapiro2009,Shapiro2020, Torrome2024},
it originates in stronger-than-classical correlations generated at the source
and accessed locally by Bob, and it persists even when the propagation channel
is entanglement breaking.

We formulate the protocol in a quantum-optical framework and treat classical
and quantum jamming on equal terms, including channel loss, imperfect idler
storage, and thermal noise. We derive the optimal cancellation strategy at
Bob's receiver and evaluate the secrecy capacity of the resulting Gaussian
wiretap channel. Correlated jamming raises that capacity by orders of
magnitude relative to unjammed transmission and enables positive secrecy even
when Eve's channel is substantially better than Bob's. We also examine the
security cost of retaining a classical record of the jamming waveform and
assess implementations in the optical and microwave domains.

The paper is organized as follows. Section~\ref{sec:protocol} introduces the
protocol and the channel model. Section~\ref{sec:cancellation} derives the
signal-to-noise ratios and the optimal noise-cancellation strategy.
Section~\ref{sec:secrecy} analyzes the secrecy capacity.
Section~\ref{sec:security} sets out the threat model and possible attacks, and
Sec.~\ref{sec:implementation} examines implementations. Section~\ref{sec:conclusion}
concludes. Detailed derivations are collected in
Appendixes~\ref{app:snr}--\ref{app:holevo}.

\begin{figure}[t!]
\centering{\includegraphics[width=1\columnwidth]{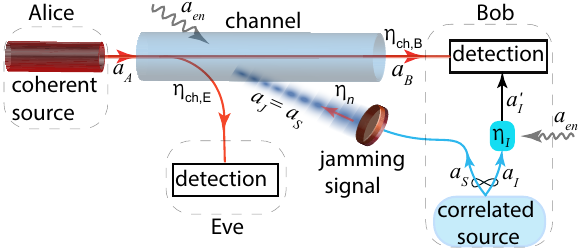}}
\caption{Schematic of the correlated-jamming protocol. Alice encodes her message in a bright coherent signal $a_{\rm A}$ that propagates to Bob over a thermal-loss channel of transmissivity $\eta_{\mathrm{ch},B}$; a passive eavesdropper Eve taps the channel with transmissivity $\eta_{\mathrm{ch},E}$. To protect the transmission, Bob operates a correlated-noise source---either a classically correlated noise generator or an entangled two-mode squeezed source---emitting a signal mode $a_{\rm S}$ and an idler mode $a_{\rm I}$ with cross correlation $\avg{a_{\rm I}a_{\rm S}}$. The signal mode is broadcast into the channel as a bright jamming field $a_{\rm J}=a_{\rm S}$, a fraction $\eta_n$ of which couples into the receiver modes, while the idler is retained at Bob's station with storage efficiency $\eta_{\rm I}$, $a_{\rm I}'=\sqrt{\eta_{\rm I}}\,a_{\rm I}+\sqrt{1-\eta_{\rm I}}\,a_{\mathrm{en}}'$, where $a_{\mathrm{en}}'$ is an environmental mode. Bob interferes his received field with the stored idler and performs homodyne detection, canceling the jamming noise from his own measurement; Eve, lacking the idler, absorbs the full jamming noise.}
\label{fig1}
\end{figure}

\section{Protocol and channel model}\label{sec:protocol}

\subsection{Physical setup}\label{subsec:setup}

The correlated-jamming setup is shown schematically in
Fig.~\ref{fig1}. Alice prepares a bright coherent state
$\vert\alpha\rangle$ in a single bosonic mode $a_{\rm A}$. The message is
encoded in the coherent amplitude, for example through Gaussian modulation of
the amplitude quadrature. For a given value of $\alpha$, the mean photon number
is
\(
N_{\rm A}=\avg{a_{\rm A}^\dagger a_{\rm A}}=|\alpha|^2.
\)
We define the quadratures as
\(
X_i=\frac{a_i+a_i^\dagger}{\sqrt{2}}, 
P_i=\frac{a_i-a_i^\dagger}{\mathrm{i}\sqrt{2}},
\)
for which the vacuum quadrature variance is $1/2$.

The signal propagates through a thermal-loss Gaussian channel.
For each party $j\in\{\mathrm{B},\mathrm{E}\}$, the received mode is
\begin{equation}
\label{eq:channel}
a_j
=
\sqrt{\eta_{\mathrm{ch},j}}\,a_{\rm A}
+
\sqrt{1-\eta_{\mathrm{ch},j}}\,a_{n,j},
\end{equation}
where $\eta_{\mathrm{ch},\mathrm{B}}$
($\eta_{\mathrm{ch},\mathrm{E}}$) shows the Alice-to-Bob
(Alice-to-Eve) transmissivity. The effective noise mode at receiver $j$ is
\begin{equation}
\label{eq:noise}
a_{n,j}
=
\sqrt{\eta_n}\,a_{{\rm J}}
+
\sqrt{1-\eta_n}\,a_{{\rm en},j},
\end{equation}
where $a_{\rm J}$ is Bob's broadcast jamming field and $a_{{\rm en},j}$ is an
environmental thermal mode with mean occupation
$\langle a_{{\rm en},j}^{\dagger}a_{{\rm en},j}\rangle=N_{\rm en}$.
The parameter $\eta_n$ is the fraction of the effective noise entering the
detected mode that originates from the jammer. We take $\eta_n$ to be the same
for Bob and Eve. This is a conservative choice that grants Bob no geometric
coupling advantage even though his receiver is co-located with the jammer and
would generally collect a larger fraction of the jamming field.

In a broadcast paltform---wireless, free space, or a guided channel into which
Bob injects his jamming field---Eqs.~(\ref{eq:channel}) and (\ref{eq:noise}) describe
the marginal channel seen by each receiver. Bob and Eve
collect different fractions of Alice's signal, while the broadcast jamming
field reaches both receivers with the same assumed coupling strength and
environmental statistics. These equations should be understood as an
effective marginal description. A simultaneous bosonic realization requires an
explicit multiport dilation with additional environmental modes and the
corresponding constraints on the channel transmissivities. A complete multiport description may be built as a
cascade of beam splitters, with Eve's tap preceding the remaining propagation
to Bob. Because the analysis below depends only on single-receiver marginal
statistics, the effective description in Eqs.~(\ref{eq:channel}) and
(\ref{eq:noise}) is sufficient; see Appendix~\ref{app:snr}. Writing
$\eta_{\rm E}\equiv\eta_n(1-\eta_{\mathrm{ch},E})$ ties the jamming Eve collects
to her own signal transmissivity, which corresponds to a collinear
Alice--Eve--Bob geometry in which an eavesdropper close to Alice is
correspondingly far from the jammer. This relation is a convenient
single-parameter geometry rather than a general free-space or guided-wave
propagation law; more general implementations require independent collection
coefficients for the Alice--Eve and Bob--Eve paths.

To protect his own receiver from the jamming field, Bob generates the field using a correlated-noise source that emits two zero-mean Gaussian modes, a signal mode $a_{\rm S}$ and an idler mode $a_{\rm I}$, with real cross-correlation $\avg{a_{\rm I}a_{\rm S}}$. The signal mode is broadcast as the jamming field,
$a_{\rm J}=a_{\rm S}$, and we write $N_{\rm J}=N_{\rm S}$ for its mean photon
number when we wish to emphasize its role as the jamming power. The idler is
retained at Bob's station, and imperfect retention---delay-line loss, for
instance---is modeled as a beam-splitter admixture of environmental noise,
\begin{equation}\label{eq:idler}
a_{\rm I}'=\sqrt{\eta_{\rm I}}\; a_{\rm I}+\sqrt{1-\eta_{\rm I}}\; a'_{\mathrm{en}} ,
\end{equation}
with storage efficiency $\eta_{\rm I}$ and
$\langle a_{\mathrm{en}}'^{\dagger}a_{\mathrm{en}}'\rangle=N_{\rm en}$. Each
temporal mode of the jamming field is paired with its own idler mode, so the
protocol runs symbol by symbol at the full modulation bandwidth of the source.

Throughout the quantitative analysis Eve is passive. She injects nothing into
the channel and collects a single spatiotemporal mode per symbol, with known or
bounded transmissivities $\eta_{\mathrm{ch},E}$ and $\eta_n$. She has access
neither to the idler nor to any record of the jamming waveform.
Section~\ref{sec:security} examines these assumptions critically.

\subsection{Classical and quantum correlated noise sources}
\label{subsec:sources}

To define a fair classical benchmark we compare two types of correlated
two-mode source. In both, the signal mode $a_{\rm J}=a_{\rm S}$ is broadcast as
the jamming field while the idler $a_{\rm I}$ is retained by Bob, and we choose
the signal modes to have the same thermal statistics and mean photon number so
that Eve observes the same local jamming field in either case.

\textbf{Classically correlated source.}
We consider any two-mode source whose state admits a nonnegative
Glauber--Sudarshan $P$ function. Such a source can be built, for
instance, by driving two optical or microwave fields from a common
electronic-noise waveform. Writing the mean photon numbers of the two modes as
\(
N_{\rm S}=\avg{a_{\rm S}^\dagger a_{\rm S}}
\) and \(
N_{\rm I}=\avg{a_{\rm I}^\dagger a_{\rm I}},
\)
the Cauchy--Schwarz inequality applied to the $P$ distribution gives the
classical bound
\begin{equation}
\label{eq:classbound}
\left|\avg{a_{\rm I}a_{\rm S}}\right|
\leq
\sqrt{N_{\rm S}N_{\rm I}}.
\end{equation}
We take the classical source to saturate this bound, and we require its signal
marginal to be phase insensitive,
\begin{equation}
\var{X_{\rm S}}=N_{\rm S}+\frac{1}{2},
\end{equation}
so that the classical and quantum jamming fields have identical local
statistics at the same $N_{\rm S}$.

\textbf{Entangled source.}
We take the quantum source to be a two-mode squeezed-vacuum state produced by
a nondegenerate parametric amplifier operated below threshold \cite{Clerk2010,Hatridge2011, zhong2013,eichler2011}. For squeezing
parameter $r$, the two modes have equal mean photon numbers \(
N_{\rm S}=N_{\rm I}=\sinh^2 r \)
and their cross correlation is
\begin{equation}
\label{eq:qbound}
\left|\avg{a_{\rm I}a_{\rm S}}\right|
=\cosh r\sinh r=
\sqrt{N_{\rm S}(N_{\rm S}+1)}
>
N_{\rm S}.
\end{equation}
At equal signal and idler brightness the classical bound in
Eq.~(\ref{eq:classbound}) is $N_{\rm S}$, so the TMSV correlation exceeds the
largest value any classical source can reach. This stronger-than-classical
correlation is the one certified by the Duan--Simon inseparability criterion
and by EPR-steering inequalities
\cite{Reid2000,Kogias2015,Uola2020}.

The correlation strength is conveniently quantified by the EPR-type joint
quadratures \(
X^-=\frac{X_{\rm S}-X_{\rm I}}{\sqrt{2}}, 
P^+=\frac{P_{\rm S}+P_{\rm I}}{\sqrt{2}}.
\) For the classical source that saturates Eq.~(\ref{eq:classbound}), their
variances are
\begin{equation}
\label{eq:varclass}
\avg{(X^-)^2}
=
\avg{(P^+)^2}
=
\frac{1}{2}+\frac{\sigma_{\rm cl}}{2}
\;\xrightarrow{N_{\rm S}=N_{\rm I}}\;
\frac{1}{2},
\end{equation}
where
\(
\sigma_{\rm cl}
=
\left(\sqrt{N_{\rm S}}-\sqrt{N_{\rm I}}\right)^2
\)
measures the mismatch between the signal and idler brightnesses. At equal
brightness the classical source reaches the vacuum level of $1/2$ but cannot
fall below it.

For the entangled source,
\begin{equation}
\label{eq:varq}
\avg{(X^-)^2}
=
\avg{(P^+)^2}
=
\frac{1}{2}+\sigma_{\rm q}
\leq
\frac{1}{2},
\end{equation}
with
\(
\sigma_{\rm q}
=
N_{\rm S}
-
\sqrt{N_{\rm S}(N_{\rm S}+1)} .
\)
For a bright source, $N_{\rm S}\gg1$,
\begin{equation}
\sigma_{\rm q}
\simeq
-\frac{1}{2}+\frac{1}{8N_{\rm S}},
\end{equation}
so the joint-quadrature variance approaches $1/(8N_{\rm S})$ and tends to zero
as the source brightens.

\section{Noise cancellation} \label{sec:cancellation}
Equations~(\ref{eq:varclass}) and (\ref{eq:varq}) capture the central idea of the protocol. Eve receives the full thermal jamming field, while Bob uses the retained idler to cancel the correlated fluctuations. Classical correlations reduce the residual noise only to the vacuum level, whereas entanglement suppresses it below this limit. In this section, we examine the conditions under which such noise cancellation can be achieved.

\subsection{Balanced receiver}
\label{subsec:balanced}

Bob's simplest receiver interferes the received mode $a_{\rm B}$ with the
stored idler $a_{\rm I}'$ on a balanced beam splitter and measures the
difference quadrature
\(
X_{\rm B}^-=(X_{\rm B}-X_{\rm I}')/\sqrt{2} .
\)
Because the idler has zero mean, this preserves Alice's signal while partially
cancelling the correlated jamming fluctuations. The resulting signal-to-noise
ratio (SNR) is derived in Appendix~\ref{app:snr},
\begin{equation}
\label{eq:SNRbob}
\snr_{\rm B}
=
\frac{\avg{X_{\rm B}^-}^2}{(\Delta X_{\rm B}^-)^2}
=
\frac{2\eta_{\mathrm{ch},B}N_{\rm A}}
{\sigma_{\mathrm{en}}^{\rm B}+\sigma_{\rm J}^{\rm B}},
\end{equation}
where
\(
\sigma_{\mathrm{en}}^{\rm B}
=
\left[
(1-\eta_{\mathrm{ch},B})(1-\eta_n)
+
1-\eta_{\rm I}
\right]N_{\mathrm{en}}
\)
collects the environmental contributions and $\sigma_{\rm J}^{\rm B}$ is the
residual jamming and vacuum noise. For the optimal classical source,
\begin{equation}
\label{eq:jamclassic}
\sigma_{\rm J}^{\rm B}
=
\eta_{\rm S}N_{\rm S}
+
\eta_{\rm I}N_{\rm I}
-
2\sqrt{\eta_{\rm S}\eta_{\rm I}N_{\rm S}N_{\rm I}}
+
1,
\end{equation}
whereas the TMSV source gives
\begin{equation}
\label{eq:jamquantum}
\sigma_{\rm J}^{\rm B}
=
(\eta_{\rm S}+\eta_{\rm I})N_{\rm S}
-
2\sqrt{\eta_{\rm S}\eta_{\rm I}
N_{\rm S}(N_{\rm S}+1)}
+
1 .
\end{equation}
Here
\(
\eta_{\rm S}
\equiv
\eta_n(1-\eta_{\mathrm{ch},B})
\)
is the effective transmissivity of the jamming field into Bob's detected mode
and $\eta_{\rm I}$ is the idler-storage efficiency. These expressions hold for
arbitrary idler loss and so describe partial cancellation; as
$\eta_{\rm I}\rightarrow0$ Bob has no useful reference and the receiver reduces
to ordinary self-jamming.

For the matched choice $\eta_{\rm I}=\eta_{\rm S}$, which can be implemented
with the tunable attenuator of Fig.~\ref{fig1}, Eqs.~(\ref{eq:jamclassic}) and
(\ref{eq:jamquantum}) reduce to
\begin{equation}
\label{eq:matched}
\sigma_{\rm J}^{\rm B}
=
\begin{cases}
\eta_{\rm S}\sigma_{\rm cl}+1,
& \text{classical},\\[2pt]
2\eta_{\rm S}\sigma_{\rm q}+1,
& \text{TMSV}.
\end{cases}
\end{equation}
where the constant $+1$ arises from the balanced receiver itself. 

For a symmetric classical source, $N_{\rm S}=N_{\rm I}$ and
$\sigma_{\rm cl}=0$, so $\sigma_{\rm J}^{\rm B}=1$: the classical excess noise
cancels completely and Bob is left with the vacuum fluctuations of the two
measured modes. For a bright TMSV source, $\sigma_{\rm q}\rightarrow-1/2$ and
therefore
\(
\sigma_{\rm J}^{\rm B}
\rightarrow
1-\eta_{\rm S}
<
1 ,
\)
so the quantum correlations suppress part of the vacuum contribution and carry
Bob below the floor attainable with classical correlations.

The condition $\eta_{\rm I}=\eta_{\rm S}$ is optimal only for the
symmetric classical source. Minimizing $\sigma_{\rm J}^{\rm B}$ over
$\eta_{\rm I}$ gives in general for a vacuum idler-loss port, $N_{\mathrm{en}}=0$
\begin{equation}
\label{eq:etaIopt}
\eta_{\rm I}^{\star}
=
\eta_{\rm S}
\frac{\avg{X_{\rm S}X_{\rm I}}^2}{N_{\rm I}^2}
=
\begin{cases}
\eta_{\rm S}N_{\rm S}/N_{\rm I},
& \text{classical},\\[2pt]
\eta_{\rm S}(N_{\rm S}+1)/N_{\rm S},
& \text{TMSV},
\end{cases}
\end{equation}
and at this optimum
\begin{equation}
\sigma_{\rm J}^{\rm B}
=
\eta_{\rm S}
\left[
N_{\rm S}
-
\frac{\avg{X_{\rm S}X_{\rm I}}^2}{N_{\rm I}}
\right]
+
1 ,
\end{equation}
which equals $1$ for any optimal classical source and $1-\eta_{\rm S}$ for a
TMSV source. Both results are exact and hold for every $N_{\rm S}$, not only in
the bright-source limit. The optimal idler transmissivity must of course satisfy
$\eta_{\rm I}^{\star}\leq 1$, so for the TMSV source the unconstrained optimum
in Eq.~(\ref{eq:etaIopt}) is physically accessible only when
\begin{equation}
\eta_{\rm S}
\leq
\frac{N_{\rm S}}{N_{\rm S}+1} ;
\end{equation}
for larger $\eta_{\rm S}$ the optimum sits at the physical boundary
$\eta_{\rm I}=1$.

Turning to Eve, she has access only to the received mode and homodynes the
quadrature carrying Alice's message. For information encoded in a single
quadrature, homodyne detection is the optimal single-mode Gaussian
measurement; heterodyne detection would split the signal between two
quadratures and add vacuum noise. Eve's SNR is therefore
\begin{equation}
\label{eq:SNReve}
\snr_{\rm E}
=
\frac{\avg{X_{\rm E}}^2}{(\Delta X_{\rm E})^2}
=
\frac{2\eta_{\mathrm{ch},E}N_{\rm A}}
{\sigma_{\mathrm{en}}^{\rm E}+\sigma_{\rm J}^{\rm E}},
\end{equation}
with environmental contribution
\(
\sigma_{\mathrm{en}}^{\rm E}
=
(1-\eta_{\mathrm{ch},E})(1-\eta_n)N_{\mathrm{en}}
\)
and combined jamming and vacuum noise
\begin{equation}
\label{eq:jameve}
\sigma_{\rm J}^{\rm E}
=
\eta_{\rm E}N_{\rm S}+\frac{1}{2} .
\end{equation}
Since the reduced state of the broadcast jamming mode is thermal for both
sources, Eve observes identical noise statistics for classical and entangled
jamming. The quantum advantage is unavailable from her local measurement and
arises entirely from Bob's access to the retained idler.

The two receivers respond differently as the jamming brightens. Eve's jamming
noise grows linearly,
$\sigma_{\rm J}^{\rm E}\simeq\eta_{\rm E}N_{\rm S}$, whereas Bob's residual
stays bounded once the idler contribution is properly adjusted, at
$\sigma_{\rm J}^{\rm B}=1$ for the optimal classical source and
$1-\eta_{\rm S}$ for the TMSV source.

For the balanced receiver, cancellation is sensitive to the idler attenuation,
because the two photocurrents are combined with fixed equal weights. The
sensitivity is quadratic: Eq.~(\ref{eq:jamclassic}) may be written exactly as
\begin{equation}
\label{eq:mismatch}
\sigma_{\rm J}^{\rm B}
=
N_{\rm I}\left(\sqrt{\eta_{\rm I}}-\sqrt{\eta_{\rm I}^{\star}}\right)^{2}
+
\sigma_{\rm J}^{\rm B}\big|_{\eta_{\rm I}^{\star}} ,
\end{equation}
and likewise for Eq.~(\ref{eq:jamquantum}). A mismatch therefore costs
$N_{\rm I}(\sqrt{\eta_{\rm I}}-\sqrt{\eta_{\rm I}^\star})^{2}$ in the noise
budget, so the tolerance tightens as the source is brightened and the balanced
receiver requires an estimate of the channel losses to set the attenuator. This
is a property of the balanced receiver rather than of the protocol, and it
disappears once the two homodyne records are combined with optimized weights.

We note in passing that $\sigma_{\rm J}^{\rm B}$ and $\sigma_{\rm J}^{\rm E}$
are referred to differently normalized signal measurements and should not be
compared directly as receiver performance metrics. The relevant condition is
$\snr_{\rm B}>\snr_{\rm E}$, which also carries the factor
$\eta_{\mathrm{ch},B}/\eta_{\mathrm{ch},E}$; this is the criterion used in
Sec.~\ref{sec:secrecy}.

\subsection{Optimal linear combining}
\label{subsec:combining}

Because homodyne detection is linear in the field quadratures, Bob need not
combine the received field and the stored idler with fixed equal weights.
Instead, he can measure $X_{\rm B}$ and $X_{\rm I}'$ separately and form the
linear estimator
\begin{equation}
\label{eq:combiner}
X_{\rm out}(g)
=
X_{\rm B}-gX_{\rm I}',
\end{equation}
choosing the electronic gain $g$ to minimize the output noise. With
\(
V_{\rm B}=\var{X_{\rm B}},
V_{\rm I}=\var{X_{\rm I}'},
C_{\rm BI}
=
\avg{X_{\rm B}X_{\rm I}'}
=
\sqrt{\eta_{\rm S}\eta_{\rm I}}\,
\avg{X_{\rm S}X_{\rm I}},
\)
the output variance is
\begin{equation}
V(g)
=
V_{\rm B}+g^2V_{\rm I}-2gC_{\rm BI} ,
\end{equation}
and minimizing over $g$ gives
\begin{equation}
\label{eq:gopt}
g^\star
=
\frac{C_{\rm BI}}{V_{\rm I}},
\qquad
V_{\min}
=
V_{\rm B}-\frac{C_{\rm BI}^2}{V_{\rm I}}.
\end{equation}

Since the idler has zero mean, the gain leaves Alice's signal untouched,
\(
\avg{X_{\rm out}(g)}^2
=
\avg{X_{\rm B}}^2
=
2\eta_{\mathrm{ch},B}N_{\rm A} ,
\)
so the optimized SNR is
\begin{equation}
\label{eq:SNRopt}
\snr_{\rm B}^{\rm opt}
=
\frac{\avg{X_{\rm out}}^2}{V_{\min}}
=
\frac{2\eta_{\mathrm{ch},B}N_{\rm A}}{V_{\min}} .
\end{equation}

Equation~(\ref{eq:SNRopt}) also reproduces the balanced-receiver result: at
$g=1$,
\begin{equation}
V(1)
=
V_{\rm B}+V_{\rm I}-2C_{\rm BI}
=
\sigma_{\mathrm{en}}^{\rm B}
+
\sigma_{\rm J}^{\rm B},
\end{equation}
and since $V(1)$ is twice the variance of
$X_{\rm B}^{-}=(X_{\rm B}-X_{\rm I}')/\sqrt{2}$, the corresponding SNR is
identical to Eq.~(\ref{eq:SNRbob}). The fixed $3$~dB penalty of the balanced
receiver appears here as the extra idler noise admitted when the two
quadratures are combined with equal weights, and the optimized receiver avoids
it by choosing $g=g^\star$. Note that because $V(g)$ is quadratic near its minimum, small
errors in the estimated gain raise the residual noise only to second order.
Electronic weighting also removes the need for a tunable idler attenuator and
so avoids the noise such an element would add. The balanced receiver is the
special case $g=1$; the optimized receiver adjusts the relative weight directly
rather than matching the two paths by physical attenuation.

\subsection{The conditional-variance bound}
\label{subsec:residual}

We now derive a source-independent criterion for whether correlated jamming
degrades, restores, or improves Bob's receiver. The same criterion identifies
the best cancellation achievable with classical correlations and the absolute
limit allowed by quantum mechanics.

We define the residual self-jamming noise as the difference between the optimized variance $V_{\min}$ and the jamming-free reference $\tfrac{1}{2}+\kappa_{\rm B}N_{\rm en}$, with \(
\kappa_{\rm B}=(1-\eta_{\mathrm{ch},B})(1-\eta_n).
\)
This reference corresponds to the same receiver when the jamming input is replaced by vacuum. For ideal idler retention, $\eta_{\rm I}=1$, the residual self-jamming noise reduces to the compact form

\begin{equation}
\label{eq:master}
\Delta V
=
\eta_{\rm S}
\left(
V_{{\rm S}|{\rm I}}-\frac{1}{2}
\right),
\end{equation}
with
\begin{equation}
\label{eq:condvardef}
V_{{\rm S}|{\rm I}}
\equiv
\var{X_{\rm S}}
-
\frac{\avg{X_{\rm S}X_{\rm I}}^2}
{\var{X_{\rm I}}} ;
\end{equation}
the derivation is given in Appendix~\ref{app:gain}. The quantity $V_{{\rm S}|{\rm I}}$ represents the minimum linear inference variance of the jamming quadrature $X_{\rm S}$ obtained from Bob's measurement of the idler quadrature $X_{\rm I}$. For jointly Gaussian states, it is equivalent to the minimum mean-square conditional variance and coincides with the inferred variance used in Reid's formulation of the EPR paradox \cite{Reid2000,Uola2020}.

Equation~(\ref{eq:master}) shows that if
$V_{{\rm S}|{\rm I}}>\tfrac{1}{2}$ the jammer increases Bob's noise; if
$V_{{\rm S}|{\rm I}}=\tfrac{1}{2}$ Bob recovers his jamming-free floor; and if
$V_{{\rm S}|{\rm I}}<\tfrac{1}{2}$ the correlated source carries him below that
floor. The source enters only through $V_{{\rm S}|{\rm I}}$, and the fraction
of the jamming field returned to Bob only through the prefactor
$\eta_{\rm S}$.

The usefulness of Eq.~(\ref{eq:master}) follows from two general bounds. For
any classically correlated source with a nonnegative Glauber--Sudarshan
$P$ function,
\begin{equation}
\label{eq:clbound}
V_{{\rm S}|{\rm I}}^{\rm cl}
\geq
\frac{1}{2} ,
\end{equation}
so classical correlations can restore Bob's receiver to the vacuum-noise floor
but cannot take it below; a derivation is given in Appendix~\ref{app:gain}. For
any physical source, on the other hand, \(V_{{\rm S}|{\rm I}}\geq0,\) which
gives the absolute quantum bound
\begin{equation}
\label{eq:qbound2}
\Delta V
\geq
-\frac{\eta_{\rm S}}{2} .
\end{equation}
The interval
\(
-\eta_{\rm S}/2
\leq
\Delta V
<
0
\)
is therefore closed to classically correlated sources and measures the
reduction available below the classical noise floor.

For the optimal classical and TMSV sources of Sec.~\ref{subsec:sources},
Eq.~(\ref{eq:master}) gives
\begin{align}
\Delta V_{\rm cl}
&=
\eta_{\rm S}
\frac{N_{\rm S}}{2N_{\rm I}+1}
\;\xrightarrow{N_{\rm I}\to\infty}\;
0,
\label{eq:residcl}\\
\Delta V_{\rm q}
&=
-\eta_{\rm S}
\frac{N_{\rm S}}{2N_{\rm S}+1}
\;\xrightarrow{N_{\rm S}\gg1}\;
-\frac{\eta_{\rm S}}{2} .
\label{eq:residq}
\end{align}
The classical residual is positive for any finite idler brightness and reaches
zero only as $N_{\rm I}\to\infty$. The TMSV residual is negative for every
$N_{\rm S}>0$ and approaches the absolute quantum bound as the source
brightens.

\begin{figure}[t!]
\centering{\includegraphics[width=1\columnwidth]{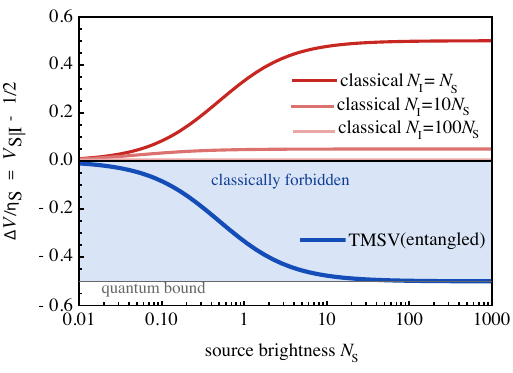}}
\caption{Residual self-jamming noise normalized to the jamming-path
transmissivity. The plotted quantity is $\Delta V/\eta_{\rm S}=V_{{\rm S}|{\rm
I}}-\tfrac12$ of Eq.~(\ref{eq:master}): positive means jamming degrades Bob,
zero that he is exactly restored to his unjammed vacuum floor, negative that he
is carried below it. No channel parameter enters, so the curves are a property
of the noise source alone. The shaded band between the classical floor (solid,
$\Delta V=0$) and the quantum bound (dashed, $\Delta V=-\eta_{\rm S}/2$) is
inaccessible to any source with a nonnegative $P$ function, however bright its
idler [Eq.~(\ref{eq:clbound})]. The symmetric classical source concedes
$+\eta_{\rm S}/2$; brightening the idler drives the classical family toward the
floor but never through it; the two-mode squeezed source lies below the floor
for every $N_{\rm S}>0$ and saturates the bound~Eq. (\ref{eq:qbound2}). The
entanglement advantage $\mathcal{A}$ of Eq.~(\ref{eq:advantage}) is the
vertical distance from the solid line to the TMSV curve. Here, $N_{\rm I}/N_{\rm S}=1$, $10$ and
$100$ for the classical family and $N_{\rm I}=N_{\rm S}$ for the TMSV with the ideal
idler retention, $\eta_{\rm I}=1$.}
\label{fig2}
\end{figure}

Figure~\ref{fig2} summarizes these results. After normalization by $\eta_{\rm S}$, the distinction between the classical and quantum regimes depends solely on the conditional variance of the source. As the jamming strength increases, classical correlations reduce the residual self-jamming noise at Bob's receiver toward zero, while quantum correlations suppress it below this classical limit and asymptotically approach the quantum bound $-\eta_{\rm S}/2$. 

Figure~\ref{fig2} also shows that a symmetric classical correlated-noise source is not optimal for jamming leading to \begin{equation}\label{eq:benchmarks}
\begin{aligned}
\Delta V_{\rm cl}^{\rm sym}\Big|_{N_{\rm I}=N_{\rm S}}
=\frac{\eta_{\rm S}N_{\rm S}}{2N_{\rm S}+1}\;\xrightarrow{N_{\rm S}\gg1}\;\frac{\eta_{\rm S}}{2}
\end{aligned}
\end{equation}
while \(\Delta V_{\rm cl}^{\rm opt}=\inf_{N_{\rm I}}\Delta V_{\rm cl}=0\). The symmetric benchmark $N_{\rm I}=N_{\rm S}$ is natural when the two outputs
of the source must be physically alike, as for two amplified ports of a common
generator or a thermal beam divided on a balanced beam splitter. The idler,
however, never enters the channel and is not charged against the jamming energy
budget, so nothing in the resource accounting forbids $N_{\rm I}\gg N_{\rm S}$.
The idler-optimized benchmark is therefore entirely legitimate, it is the one a
determined classical jammer would adopt, and it reaches the vacuum floor
exactly. We take $\Delta V_{\rm cl}^{\rm opt}=0$ as the classical reference
throughout and quote the entanglement advantage as
\begin{equation}\label{eq:advantage}
\begin{aligned}
\mathcal{A}&\equiv\Delta V_{\rm cl}^{\rm opt}-\Delta V_{\rm q}\\[2pt]
&=\eta_{\rm S}\,\frac{N_{\rm S}}{2N_{\rm S}+1}\;\xrightarrow{N_{\rm S}\gg1}\;\frac{\eta_{\rm S}}{2} ,
\end{aligned}
\end{equation}
which saturates the universal bound~(\ref{eq:qbound2}) and is already within a
factor of two of its asymptote at $N_{\rm S}\sim1$. Against the symmetric
classical source one would instead find $\Delta V_{\rm cl}^{\rm
sym}-\Delta V_{\rm q}=2\eta_{\rm S}N_{\rm S}/(2N_{\rm S}+1)\to\eta_{\rm S}$,
twice as large. 

The Eq.~(\ref{eq:advantage}) shows the central
quantum-optical mechanism of the protocol. By optimally combining $X_{\rm B}$
with the idler of a two-mode squeezed-vacuum state, Bob performs a conditional
inference measurement that reduces the variance of the jamming quadrature below
the positive-$P$ classical limit. A single sub-vacuum conditional variance
establishes this nonclassical advantage but is not, by itself, a complete
EPR-steering test. For the phase-symmetric TMSV source, the conjugate
conditional variance obeys the same relation, and the product
$V_{X_{\rm S}|X_{\rm I}}V_{P_{\rm S}|P_{\rm I}}<1/4$ satisfies Reid's
two-quadrature steering criterion \cite{Reid2000,Kogias2015,Uola2020}. The
retained idler therefore produces sub-vacuum conditional noise in Bob's
estimator; for the electronically optimized receiver, this should not be
interpreted as a freely propagating squeezed output mode.

This advantage does not allow the jamming power to increase indefinitely without penalty, since $\mathcal{A}$ saturates rather than growing without bound. Instead, entanglement converts the vacuum floor from a fundamental limit into a resource. 

\subsection{Imperfect idler storage}
Equation~(\ref{eq:master}) assumes ideal retention. For a general storage
efficiency $\eta_{\rm I}$, the same optimal-combining analysis gives
\begin{equation}\label{imp}
\Delta V
=
\eta_{\rm S}
\left(
V_{{\rm S}|{\rm I}'}
-
\frac{1}{2}
\right),
\end{equation}
where $X_{\rm I}'$ is the attenuated idler quadrature. The classical and
quantum bounds therefore stand, although idler loss changes how closely each
source approaches them.

\begin{figure}[t!]
\centering{\includegraphics[width=0.92\columnwidth]{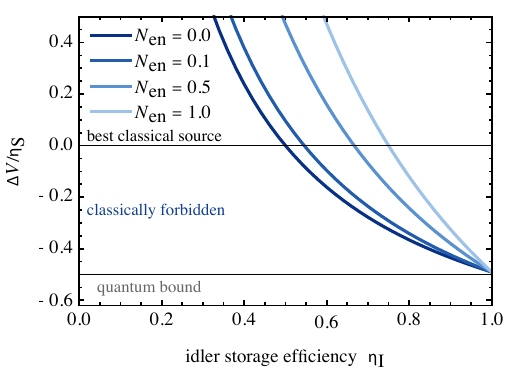}}
\caption{ Effect of imperfect idler storage on the entangled source, from
Eq.~(\ref{imp}) at $N_{\rm S}=N_{\rm I}=25$, for storage-line
occupations $N_{\mathrm{en}}=0$, $0.1$, $0.5$ and $1$. Each curve crosses the
classical floor at the threshold $\eta_{\rm I}^{\rm th}$ of
Eq.~(\ref{eq:etaIthresh}).
The threshold is
independent of $N_{\rm S}$ and of $\eta_{\rm S}$: better than $50\%$ storage on
a cold line, tightening toward unity as the line warms. The best classical
source sits on the floor for every $\eta_{\rm I}$, since a brighter idler
compensates any fixed attenuation, so idler loss penalizes the entangled source
alone. }
\label{fig3}
\end{figure}

Figure~\ref{fig3} shows the effect on cancellation, and the two sources
respond quite differently. For any nonzero storage efficiency an optimized
classical source still approaches $\Delta V_{\rm cl}=0$ in the limit of a
sufficiently bright idler, because increased idler brightness compensates for
fixed attenuation. No such compensation exists for the entangled source, where
reduction below the classical floor, $\Delta V_{\rm q}<0$, requires
(Appendix~\ref{app:gain})
\begin{equation}
\label{eq:etaIthresh}
\eta_{\rm I}
>
\eta_{\rm I}^{\rm th}
=
\frac{2N_{\mathrm{en}}+1}{2N_{\mathrm{en}}+2} .
\end{equation}
This threshold is independent of both the source brightness $N_{\rm S}$ and the
jamming return efficiency $\eta_{\rm S}$. For a vacuum idler environment,
$N_{\mathrm{en}}=0$, more than half the idler must be retained; thermal
occupation of the storage line drives the requirement toward unity.

Above the threshold, the bright-source quantum advantage falls from its ideal
value $\eta_{\rm S}/2$ to
\begin{equation}
\label{eq:advloss}
\mathcal{A}(\eta_{\rm I})
=
\frac{\eta_{\rm S}}{2}
\left[
1-
\frac{(1-\eta_{\rm I})(2N_{\mathrm{en}}+1)}
{\eta_{\rm I}}
\right] .
\end{equation}
Idler loss thus penalizes the entangled source directly, while an optimized
classical source absorbs the same attenuation by brightening its retained
reference. The storage efficiency $\eta_{\rm I}$ is accordingly the central
experimental requirement of the protocol, with Eq.~(\ref{eq:etaIthresh}) fixing
the minimum value needed to pass the classical floor.

\section{Secrecy analysis}\label{sec:secrecy}

\subsection{Secrecy capacity of the induced wiretap channel}
\label{subsec:capacity}

After Bob applies noise cancellation, the system formed by Alice, Bob, and Eve
is a Gaussian wiretap channel \cite{Wyner1975,LeungYanCheong1978} whose
performance is fixed by the two SNRs derived above. For Gaussian signaling and
homodyne detection at both receivers, the achievable secrecy rate per channel
use is \cite{LeungYanCheong1978,ElGamal2011}
\begin{equation}
\label{eq:secrecy_cap_SNR}
C
=
\max\!\left[
0,\,
\frac{1}{2}
\log_2\!\left(
\frac{1+\snr_{\rm B}}
{1+\snr_{\rm E}}
\right)
\right].
\end{equation}
For secure message transmission, $C$ is the highest rate at which Alice can
communicate reliably with Bob while Eve's leakage rate vanishes asymptotically. When
the same channel is used for secret-key generation with public discussion, the
expression is also a lower bound on the achievable key rate
\cite{Ahlswede1993,Maurer1993}.

The secrecy capacity of a wiretap channel is determined solely by the marginal conditional distributions $p(y|x)$ and $p(z|x)$ \cite{Csiszar1978}. Correlations between the noise observed by Bob and Eve, which arise because both receivers are exposed to the same jamming field, therefore do not enter Eq.~(\ref{eq:secrecy_cap_SNR}) explicitly. This supports the use of Eqs.~(\ref{eq:channel})--(\ref{eq:noise}) for marginal rate calculations, but those equations should not be read as a complete joint bosonic dilation. Appendix~\ref{app:wiretap} examines this point in more detail, including the degraded nature of Eve's jammed channel, the correlations between the receiver noises, and the measurement available to Eve. It further shows that retained-idler jamming increases the secrecy rate, $C'\geq C$, whenever Bob's residual self-jamming penalty is smaller than the additional noise imposed on Eve.

Equation~(\ref{eq:secrecy_cap_SNR}) assumes homodyne detection at Eve. To assess security against a more capable adversary, Appendix~\ref{app:holevo} evaluates Eve's Holevo information $\chi_{\rm E}$ under optimal collective measurements on her Gaussian output states. A data-processing argument shows that jamming reduces $\chi_{\rm E}$ monotonically. In the strong-jamming regime, the Holevo and homodyne descriptions converge, with their relative difference scaling as $1/(12\bar n_{\rm E}^{2})$, where
\begin{equation}
\bar n_{\rm E}=\eta_{\rm E}N_{\rm S}
+
\kappa_{\rm E}N_{\mathrm{en}}
\end{equation}
is the total noise occupation received by Eve. 

\begin{figure}[t!]
\centering{\includegraphics[width=1\columnwidth]{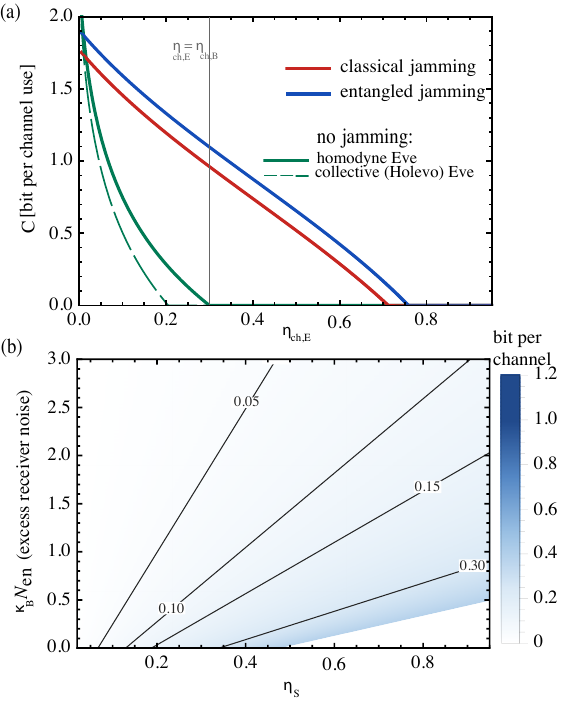}}
\caption{ (a) Secrecy capacity $C$ versus Eve's channel transmissivity
$\eta_{\mathrm{ch},E}$, without jamming (green) and with jamming by the
symmetric classical (red) and entangled (blue) sources. Solid curves: homodyne
Eve, Eq.~(\ref{eq:secrecy_cap_SNR}). Dashed curves: collective Eve,
Eq.~(\ref{eq:CHol}), with $\chi_{\rm E}$ from Appendix~\ref{app:holevo}. The dashed curves are achievable collective-attack rates and lower bounds on the secrecy capacity.
Without jamming, secrecy is lost as soon as Eve's channel rivals Bob's, at
$\eta_{\mathrm{ch},E}=0.300$ against a homodyne eavesdropper and already at
$0.205$ against a collective one. Correlated jamming moves the cutoff to
$0.712$ (classical) and $0.759$ (entangled), and simultaneously closes the gap
between the two adversary models: the corresponding collective-attack cutoffs
are $0.710$ and $0.755$, below the line width. Parameters are $N_{\rm J}=N_{\rm
S}=N_{\rm I}=25$, $N_{\rm A}=20$, $N_{\mathrm{en}}=0.1$,
$\eta_{\mathrm{ch},B}=0.3$ and $\eta_n=0.3$, giving $\eta_{\rm S}=0.21$ and
$\kappa_{\rm B}=0.49$. Bob uses the matched balanced receiver, for which
$\tfrac12\log_2(1+\snr_{\rm B})=2.257$, $1.770$ and $1.906$~bit per channel use
without jamming, with classical jamming and with entangled jamming
respectively. (b) The entanglement gain $\mathcal{G}$ of Eq.~(\ref{eq:gain}), in bits
per channel use, over the two parameters that control it: the fraction
$\eta_{\rm S}$ of the jamming field returning to Bob and his excess receiver noise $\kappa_{\rm B}N_{\mathrm{en}}$. 
}
\label{fig4}
\end{figure}

It is useful to convert
Eq.~(\ref{eq:advantage}) into a receiver specification. With optimal combining
the best classical jammer sits exactly at Bob's unjammed floor
$\tfrac12+\kappa_{\rm B}N_{\mathrm{en}}$, and a bright entangled one sits
$\eta_{\rm S}/2$ below it. Since $\snr_{\rm E}$ cancels between the two
capacities, their difference at high $\snr_{\rm B}$ is just the ratio of these
two noise budgets,
\begin{equation}\label{eq:gain}
\mathcal{G}\equiv C_{\rm q}-C_{\rm cl}
\approx\frac12\log_2\frac{1+2\kappa_{\rm B}N_{\mathrm{en}}}
                    {1+2\kappa_{\rm B}N_{\mathrm{en}}-\eta_{\rm S}} .
\end{equation}
Neither the signal brightness $N_{\rm A}$ nor the jamming brightness
$N_{\rm J}$ appears. In this limit the entanglement advantage is a fixed
improvement in Bob's receiver noise figure---$0.92$~dB, or $0.153$~bit per
channel use, at the parameters used below---and it does not fade as the signal
is brightened, because brightening the signal adds no noise to Bob's floor. Two
things erode it. One is a small $\eta_{\rm S}$, so that little of the jamming
returns to Bob to be cancelled. The other is a thermal-noise-dominated
receiver, $\kappa_{\rm B}N_{\mathrm{en}}\gg\tfrac12$, in which the fixed
$\eta_{\rm S}/2$ that entanglement removes is a negligible part of the budget. Figure~\ref{fig4}(b) maps $\mathcal{G}$ over these two parameters. Any excess noise referred to Bob's input---finite homodyne efficiency in particular, which contributes $(1-\eta_d)/2\eta_d$ for detection efficiency $\eta_d$---enters on the same axis as $\kappa_{\rm B}N_{\mathrm{en}}$ and erodes $\mathcal{G}$ in the same way.

\subsection{Jamming-enabled secrecy}\label{subsec:enabled}

Figure~\ref{fig4} shows the secrecy capacity $C$ as a function of Eve's channel transmissivity $\eta_{\mathrm{ch},E}$, which serves as a proxy for her distance from the transmitter in any geometry, guided or unguided, in which collection efficiency falls with separation. The limit $\eta_{\mathrm{ch},E}\to1$ corresponds to Eve approaching the transmitter, whereas $\eta_{\mathrm{ch},E}\to0$ corresponds to a distant eavesdropper. The main result is not merely an increase in capacity, but a substantial extension of the region over which secure communication remains possible.

In the absence of jamming, positive secrecy requires $\eta_{\mathrm{ch},E}<\eta_{\mathrm{ch},B}$, which is the standard wiretap condition requiring Eve's channel to be weaker than Bob's. Secrecy therefore vanishes as Eve moves sufficiently close to the transmitter. With correlated jamming, Eve's noise increases through the contribution $\eta_{\rm E}N_{\rm J}$ in Eq.~(\ref{eq:SNReve}), while the matched balanced receiver used in Fig.~\ref{fig4} keeps Bob's residual noise bounded. As a result, the condition $\snr_{\rm B}>\snr_{\rm E}$ persists well beyond $\eta_{\mathrm{ch},E}=\eta_{\mathrm{ch},B}$. For the parameters used in Fig.~\ref{fig4}, the secure region expands from $\eta_{\mathrm{ch},E}\lesssim0.30$ without jamming to approximately $\eta_{\mathrm{ch},E}\simeq0.71$ with correlated jamming.

The dashed curves show that this extension persists even when Eve is allowed an optimal collective measurement. Without jamming, such a measurement eliminates secrecy already at $\eta_{\mathrm{ch},E}=0.205$. Correlated jamming therefore converts channels with zero secrecy capacity into channels with positive secrecy capacity. This function resembles relay-assisted cooperative jamming in classical physical-layer security \cite{GoelNegi2008,Tekin2008,Dong2009}, but avoids the self-jamming penalty that limits those schemes. An entangled source provides an additional quantum-enhanced margin by reducing Bob's residual noise below the classical limit.

\subsection{Range and the eavesdropper exclusion zone}
\label{subsec:range}

Once a propagation model is specified, the transmissivity axis in Fig.~\ref{fig4} can be converted into distance, providing a more direct operational interpretation of the results. The conversions below are illustrative and use the single-parameter relation $\eta_{\rm E}=\eta_n(1-\eta_{\mathrm{ch},E})$. A general wireless, free-space, or guided-wave geometry requires independent path losses for Alice's signal and Bob's jamming field.

Unlike QKD, the present protocol does not require a fragile quantum state carrying the message to survive the Alice--Bob channel. Alice transmits a bright classical field that, in principle, can be amplified or regenerated. The communication range is therefore determined by the conventional link budget rather than by the repeaterless rate--loss bound \cite{Pirandola2017}. Correlated jamming does not directly increase the maximum Alice--Bob range. Instead, it reduces the exclusion region around Alice within which an eavesdropper would otherwise prevent secure communication.

For a wireless or free-space link in the far field, the transmissivity scales as $\eta\propto d^{-2}$, giving $d\propto\eta^{-1/2}$. Without jamming, positive secrecy requires Eve to be farther from Alice than Bob, so the entire disk of radius $d_{\rm B}$ around Alice must be physically protected. For the parameters used in Fig.~\ref{fig4}, correlated jamming shifts the cutoff in Eve's channel transmissivity from $\eta_{\mathrm{ch},E}=0.300$ to $0.712$ for the classical source and to $0.759$ for the entangled source. These values reduce Eve's minimum secure standoff distance to $0.649\,d_{\rm B}$ and $0.629\,d_{\rm B}$, respectively. The corresponding area requiring physical protection is reduced by factors of approximately $2.4$ and $2.5$. Equivalently, for a fixed Eve standoff distance, Bob can be located approximately $54\%$ farther from Alice with classical correlated jamming and $59\%$ farther with entangled jamming while maintaining positive secrecy.

For an optical fiber with attenuation $0.2$~dB/km, $\eta_{\mathrm{ch},B}=0.30$ corresponds to an Alice--Bob separation of approximately $26$~km. Without jamming, an eavesdropper who fully intercepts the transmitted field must be located beyond Bob for the secrecy capacity to remain positive. Classical correlated jamming reduces Eve's minimum secure distance from Alice to $7.4$~km, while entangled jamming reduces it further to $6.0$~km. An interception point between $7.4$ and $26$~km, which would eliminate secrecy without jamming, therefore becomes tolerable with classical correlated jamming. The entangled source extends this secure region further, allowing interception points beyond $6.0$~km.

\subsection{Absolute capacities and the role of jamming brightness}
\label{subsec:brightness}
\begin{figure}[t!]
\centering
\includegraphics[width=0.92\columnwidth]{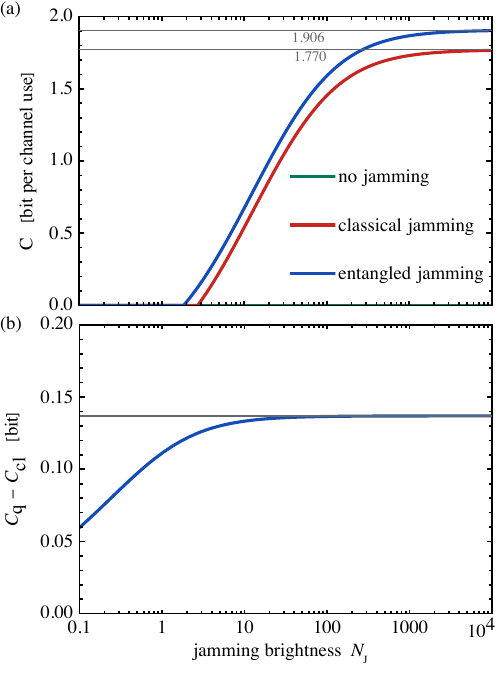}
\caption{
(a) Secrecy capacities as functions of the jamming brightness $N_{\rm J}$ for $\eta_{\mathrm{ch},E}=0.299$, just inside the unjammed cutoff $\eta_{\mathrm{ch},B}=0.300$. The unjammed capacity remains fixed at $C_{\mathrm{no\mbox{-}jamm}}=2.39\times10^{-3}$~bit per channel use. At $N_{\rm J}=25$, the classical and quantum capacities are $0.96$ and $1.10$~bit per channel use and saturate at $1.770$ and $1.906$~bit per channel use, respectively.
(b) Quantum contribution $C_{\rm q}-C_{\rm cl}$, which saturates at $0.137$~bit per channel use, consistent with $\mathcal{A}\rightarrow\eta_{\rm S}/2$. The high-$\snr_{\rm B}$ asymptotic value for this receiver is $0.149$~bit per channel use. Parameters are $N_{\rm A}=20$, $N_{\mathrm{en}}=0.1$, and $\eta_n=0.3$; all other parameters are the same as in Fig.~\ref{fig4}.
}
\label{fig5}
\end{figure}

Figure~\ref{fig5}(a) shows the secrecy capacity as the jamming brightness increases, with Eve positioned just inside the unjammed secrecy cutoff. The unjammed capacity remains fixed at $2.4\times10^{-3}$~bit per channel use, while the classical and quantum capacities rise rapidly near $N_{\rm J}\sim10$ and saturate at approximately $1.77$ and $1.91$~bit per channel use, respectively, corresponding to an enhancement of nearly three orders of magnitude.

Figure~\ref{fig5}(b) shows the quantum contribution $C_{\rm q}-C_{\rm cl}$. The gain increases with $N_{\rm J}$ and saturates near $0.137$~bit per channel use, consistent with $\mathcal{A}\rightarrow\eta_{\rm S}/2$. Entanglement therefore retains a finite advantage in the strong-jamming regime. Once the system enters the strong-jamming regime, further increases in $N_{\rm J}$ do not change the separation between the classical and quantum residual noise floors. In the regime of Eq.~(\ref{eq:gain}), this separation depends only on $\eta_{\rm S}$ and $\kappa_{\rm B}N_{\mathrm{en}}$, rather than on $N_{\rm A}$ or $N_{\rm J}$.

The practical value of this fixed quantum contribution depends on the operating regime. When the secrecy capacity is already large, as in bright radio-frequency or classical optical links, the quantum correction forms only a small fraction of the total rate. Classically correlated jamming then captures most of the benefit using an arbitrary-waveform generator and a stored record of the transmitted noise.

Entanglement becomes more important when the secrecy capacity is small or receiver noise limits performance, as in quantum-limited optical links, cryogenic microwave networks, and covert or power-constrained communication. The difference is largest near the secrecy cutoff, where a modest absolute increase produces a large relative gain. At $\eta_{\mathrm{ch},E}=0.7$ in Fig.~\ref{fig4}, the entangled source provides about $0.17$~bit per channel use, compared with $0.03$~bit per channel use for the classical source. The quantum advantage then corresponds to suppressing Bob's residual noise by $\eta_{\rm S}/2$ below the vacuum-noise floor reached by the optimal classical source.

\section{Security discussion}\label{sec:security}

\subsection{Threat model and assumptions}
\label{subsec:threat}

The security guarantees derived here follow the physical-layer wiretap model rather than the unrestricted adversarial model used in QKD. The analysis relies on the following assumptions.

\textbf{(i) Bounded collection.}
Eve's collection efficiencies for Alice's signal and Bob's jamming field, $\eta_{\mathrm{ch},E}$ and $\eta_{\rm E}$, are assumed to be known or bounded from above. Such bounds may arise from an exclusion zone, antenna directivity, spatial filtering, or monitoring of the channel-loss budget. Figure~\ref{fig4} therefore represents a worst-case analysis within these collection limits.

\textbf{(ii) Passive eavesdropping.}
Eve observes the transmitted field but does not inject signals into the channel. An active adversary could jam Bob, impersonate Alice, or modify the channel parameters. Protection against such attacks requires authentication and channel monitoring, which fall outside the physical-layer confidentiality mechanism considered here.

\textbf{(iii) Measurement model.}
The SNR analysis in the main text assumes that Eve performs homodyne detection of the quadrature carrying Alice's message. Appendix~\ref{app:holevo} considers a stronger adversary by evaluating Eve's Holevo information $\chi_{\rm E}$ under collective measurements on the Gaussian channel. In the strong-jamming regime, the difference between the homodyne and Holevo descriptions decreases as $1/\bar n_{\rm E}^{2}$. The present treatment does not constitute a general security proof against arbitrary coherent attacks across multiple channel uses.

\textbf{(iv) No access to the idler.}
The retained idler remains inside Bob's station and is inaccessible to Eve. Leakage from the idler path would weaken Bob's correlation advantage and must be included in an implementation-level security analysis.

Under these assumptions, the protocol provides information-theoretic secrecy in the wiretap sense \cite{Wyner1975,Bloch2011}. It does not adopt the assumption-free channel model of QKD, in which Eve is granted access to a purification of the communication channel. The two approaches therefore address different operating regimes. QKD uses a stronger adversarial model but requires the transmission of fragile quantum states and is constrained by rate--loss bounds \cite{Pirandola2017,Diamanti2016}. Correlated jamming instead retains bright classical signaling and uses a local quantum resource, at the cost of assuming bounded collection and a protected idler.

\subsection{The no-record advantage of entangled jamming}
\label{subsec:norecord}

The entangled source provides a second advantage beyond reducing Bob's residual noise. Optimal classical cancellation requires Bob to retain a reference strongly correlated with the transmitted jamming waveform. Depending on the implementation, this reference may consist of a stored waveform, a pseudorandom seed, or a recorded analog-noise trace. An adversary who gains access to this information before its secure deletion could use it to estimate and subtract the jamming contribution from previously recorded measurements.

The entangled source requires no classical copy of the transmitted noise. Bob instead uses the retained idler mode as a reference and measures it during reception. Provided that the idler remains secure and that no classical measurement record contains enough information to reconstruct the transmitted jamming realization, compromising Bob's control electronics does not by itself reveal an exact copy of the jamming waveform. The entangled scheme therefore removes the need for a pre-existing exact waveform or seed. The idler photocurrent is nevertheless security-sensitive side information and should also be protected and erased after use.

The conditional-variance analysis of Sec.~\ref{subsec:residual} quantifies this distinction. For the optimized classical source,
\begin{equation}
V_{{\rm S}|{\rm I}}^{\rm cl}=
\frac{1}{2}
+
\frac{N_{\rm S}}{2N_{\rm I}+1},
\end{equation}
so the vacuum floor is approached only when Bob retains a bright, high-quality classical reference. Better classical cancellation therefore requires a more accurate description of the transmitted noise, which also makes this information more useful to an adversary. The entangled source avoids this trade-off. Its retained idler provides a stronger-than-classical reference without requiring a stored waveform and allows the residual noise to fall below the classical limit in Eq.~(\ref{eq:qbound2}).

\subsection{Attack scenarios}
\label{subsec:attacks}

\textbf{Multimode collection and taps on the jamming path.}
The main analysis assigns Eve one detected mode per channel use. A stronger adversary could collect Alice's signal and Bob's jamming radiation through separate apertures and combine the resulting measurements. This is the most significant attack considered here. In a wireless or free-space geometry, the jammer acts as a localized source, and its contribution may occupy a spatial mode that Eve could partially suppress using multi-aperture beamforming. Eve would still lack access to Bob's retained idler, but the achievable suppression would depend on the spatial-mode structure, correlations, and collection efficiencies available to her.

A complete analysis of this attack requires a multimode channel model and lies beyond the present single-mode treatment. The bounded-collection assumption must therefore apply to the total collection efficiency across all modes accessible to Eve. The wireless estimates in Sec.~\ref{subsec:range} should likewise be interpreted as applying to a single-aperture eavesdropper. Guided channels, including optical fibers and cryogenic microwave transmission lines, are less exposed to this attack because the waveguide constrains the modes available to Eve.

\textbf{Channel-parameter estimation.}
The balanced receiver requires the idler transmissivity to remain close to the optimal value $\eta_{\rm I}^{\star}$ given in Eq.~(\ref{eq:etaIopt}). Variations in the jamming return path therefore degrade cancellation according to Eq.~(\ref{eq:mismatch}). The optimal-combining receiver is less sensitive because the gain $g^{\star}$ can be estimated from the locally measured covariance between Bob's signal and idler photocurrents and updated during operation. Independent channel monitoring remains necessary to detect deliberate changes to the propagation path.

\textbf{Denial of service.}
An active adversary can reduce the communication rate by injecting strong noise or blocking the channel. The protocol protects confidentiality under the stated threat model, but it does not guarantee channel availability.

\section{Experimental feasibility}
\label{sec:implementation}

\textbf{Optical domain.}
The entangled source may be realized using a below-threshold optical parametric oscillator or amplifier that generates two-mode squeezed vacuum. Two-mode squeezing above 10~dB has been demonstrated \cite{Eberle2013}, while single-mode squeezing approaching 15~dB has been reported \cite{Vahlbruch2016}.
The value $N_{\rm S}=25$ used in Figs.~\ref{fig4} and \ref{fig5} corresponds to approximately 20~dB of ideal two-mode squeezing and exceeds the performance of most current sources. We use this value to illustrate the asymptotic behavior rather than as an immediate experimental target. Equations~(\ref{eq:jamclassic})--(\ref{eq:jamquantum}) also apply at lower brightness, where presently available squeezing levels still permit sub-vacuum cancellation, although with a smaller quantum advantage.

The idler may be retained using an optical delay line. Fiber loss near $0.2$~dB/km is compatible with microsecond-scale delays, which are sufficient when the idler must be stored only for the round-trip propagation time of the jamming field. Longer delays would require an optical quantum memory \cite{Lvovsky2009,Lei2023}. Equation~(\ref{eq:etaIthresh}) sets the efficiency requirement. For a vacuum storage environment, $N_{\rm en}=0$, the combined idler-storage and detection efficiency must exceed $50\%$ for the residual noise to fall below the classical limit. Retaining most of the ideal quantum advantage requires a considerably higher efficiency. The receiver must also maintain phase coherence between the retained idler and the returned jamming field, with stability comparable to that required for continuous-variable homodyne detection. Temporal-mode mismatch and a phase error $\phi$ reduce the useful covariance, approximately by a factor $\cos\phi$ for a pure phase offset. Electronic subtraction also occurs after photodetection, so detector saturation and front-end dynamic range remain practical constraints

\textbf{Microwave domain.}
Superconducting parametric amplifiers and converters generate propagating two-mode squeezed microwave fields with the required correlations \cite{zhong2013,eichler2011,castellanos2008, Abdul24}. Such fields have been employed in microwave quantum-illumination and noise-radar experiments \cite{Barzanjeh2015,Barzanjeh2020,Luong2020,Assouly2023}, as well as in secure microwave state transfer \cite{Pogorzalek2019,Casariego2023}. The main limitations are transmission loss between the cryogenic source and receiver, amplifier noise, and thermal occupation of the idler-storage path. These effects enter the model through $\eta_{\rm I}$, $\eta_n$, and $N_{\rm en}$.
Coherent microwave links between superconducting circuits housed in separate
dilution refrigerators have been demonstrated over several meters \cite{Magnard2020},
which is precisely the setting envisaged for a first demonstration of the protocol.

The microwave implementation is especially sensitive to thermal noise. A warm idler path increases $N_{\rm en}$ and shifts the threshold $\eta_{\rm I}^{\rm th}$ in Eq.~(\ref{eq:etaIthresh}) toward unity. The source, idler-storage path, and first-stage receiver should therefore remain cryogenic. Short links between cryogenic devices provide the most practical setting for an initial demonstration.

\textbf{Classical implementation as a deployment path.}
A classical implementation may use an arbitrary-waveform generator to produce broadband Gaussian noise while retaining a digital copy of the transmitted waveform. Bob then combines this reference with the received signal using the optimal electronic receiver defined in Eq.~(\ref{eq:combiner}). This approach follows the self-interference-cancellation architecture developed for in-band full-duplex radio \cite{Zheng2013,Sabharwal2014}. After calibration of the transmitter and propagation path, the stored waveform provides a high-quality estimate of the jamming field reaching Bob.

For a bright and nearly noiseless reference, the classical residual approaches \(
\Delta V_{\rm cl}\rightarrow0 \)
and Bob recovers the unjammed vacuum-noise floor. This represents the strongest classical benchmark considered in this work. Reaching this limit requires electronic combining rather than passive optical attenuation, since the reference photocurrent must be assigned an independently optimized weight. In practical radio-frequency hardware, transmitter phase noise, amplifier nonlinearities, and converter quantization usually set the cancellation floor before vacuum fluctuations become relevant. The sub-vacuum quantum margin is therefore most relevant in shot-noise-limited systems.

\section{Conclusions}
\label{sec:conclusion}

We have introduced a correlated-jamming protocol that addresses the main limitation of cooperative jamming, namely self-interference at the legitimate receiver. Alice transmits a bright coherent signal, while Bob broadcasts one mode of a correlated two-mode source and retains the second as a local reference. Eve receives the full thermal jamming field, whereas Bob uses the retained idler to suppress the correlated fluctuations in his measurement.

The residual self-jamming noise is governed by the conditional variance of the transmitted jamming quadrature inferred from the retained idler. This links the cancellation mechanism to EPR quadrature inference and, for a phase-symmetric TMSV source, to the corresponding two-quadrature steering criterion. Classical correlations reduce Bob's residual noise to the vacuum floor but cannot push it below this limit. Entangled two-mode squeezed states surpass the classical bound and produce sub-vacuum residual noise, yielding a finite advantage over the optimal classical strategy.

Both classical and quantum correlated jamming strongly reduce Eve's signal-to-noise ratio and increase the secrecy rate of the resulting Gaussian wiretap channel. Positive secrecy therefore persists even when Eve has a stronger direct channel than Bob. Correlated jamming provides most of this improvement, while entanglement adds a quantum margin by lowering Bob's noise below the classical floor. In the strong-jamming regime, this margin is fixed by the jamming return efficiency and Bob's receiver noise rather than by the signal or jamming power.

The quantum advantage depends critically on idler retention. Loss, detector inefficiency, and thermal occupation weaken the available correlations. A classical source tolerates fixed reference loss more readily by increasing the brightness and quality of the retained reference, but this requires storing a more accurate record of the transmitted jamming waveform. The entangled protocol avoids the need for a pre-existing exact stored waveform by using a protected quantum reference measured locally during reception, although the resulting idler record remains sensitive.

Potential applications of our proposal include secure wireless, fiber-optic, and free-space links with enforced signal--jammer mode overlap, shot-noise-limited optical communication, cryogenic microwave networks, and covert or power-constrained communication. A classical implementation using a broadband noise source and adaptive electronic cancellation offers a direct route toward experimental validation, while short optical delay lines and cryogenic microwave links provide suitable platforms for testing the quantum version.

This approach sits between conventional physical-layer security and QKD. It preserves bright classical signaling and avoids transmitting a fragile message-bearing quantum state, but relies on bounded collection by Eve, passive eavesdropping, and a protected idler path. Future work should address coherent attacks across multiple channel uses, multimode collection, active channel manipulation, imperfect parameter estimation, and experimental validation under realistic loss and noise.

\begin{acknowledgments}
S.B. thanks Stefano Pirandola for helpful comments. S.B. and B.C.S.\ acknowledge funding from the Natural Sciences and Engineering Research Council of Canada (NSERC) through its Discovery Grant program, and support from the University of Calgary.
\end{acknowledgments}

\appendix

\section{Derivation of the measurement signal-to-noise ratios}\label{app:snr}

All states are zero-mean Gaussian apart from Alice's coherent displacement, so second moments suffice. With the convention $X=(a+a^\dagger)/\sqrt2$ (vacuum variance $1/2$), the relevant input moments are as follows. Alice's coherent signal has $\avg{X_{\rm A}}=\sqrt{2N_{\rm A}}$ and $\var{X_{\rm A}}=1/2$; each environmental mode has $\var{X_{\mathrm{en}}}=N_{\mathrm{en}}+1/2$ (every occurrence of $a_{\mathrm{en}}$ denotes an independent reservoir mode of the same occupation); and the source quadratures have $\var{X_{\rm S}}=N_{\rm S}+\tfrac12$ and $\var{X_{\rm I}}=N_{\rm I}+\tfrac12$, with cross correlation
\begin{equation}
\avg{X_{\rm S}X_{\rm I}}=
\begin{cases}
\sqrt{N_{\rm S}N_{\rm I}} & \text{(optimal classical)},\\[2pt]
\sqrt{N_{\rm S}(N_{\rm S}+1)} & \text{(TMSV)} .
\end{cases}
\end{equation}

From Eqs.~(\ref{eq:channel})--(\ref{eq:idler}),
\begin{align}
\var{X_{\rm B}}&=\frac{\eta_{\mathrm{ch},B}}{2}+\eta_{\rm S}\Big(N_{\rm S}+\frac12\Big)+\kappa_B\Big(N_{\mathrm{en}}+\frac12\Big),\label{eq:VB}\\
\var{X_{\rm I}'}&=\eta_{\rm I}\Big(N_{\rm I}+\frac12\Big)+(1-\eta_{\rm I})\Big(N_{\mathrm{en}}+\frac12\Big),\label{eq:VI}\\
\avg{X_{\rm B}X_{\rm I}'}&=\sqrt{\eta_{\rm S}\eta_{\rm I}}\;\avg{X_{\rm S}X_{\rm I}},\label{eq:CBI}
\end{align}
with $\eta_{\rm S}=\eta_n(1-\eta_{\mathrm{ch},B})$ and $\kappa_B=(1-\eta_{\mathrm{ch},B})(1-\eta_n)$. The vacuum contributions in Eqs.~(\ref{eq:VB}) and (\ref{eq:VI}) sum to $1/2$ each, since $\eta_{\mathrm{ch},B}+\eta_{\rm S}+\kappa_B=1$ and $\eta_{\rm I}+(1-\eta_{\rm I})=1$, confirming that the passive network preserves the commutation relations.

For the balanced combination $X_{\rm B}^-=(X_{\rm B}-X_{\rm I}')/\sqrt2$,
\begin{align}
\avg{X_{\rm B}^-}^2&=\eta_{\mathrm{ch},B}N_{\rm A},\\
(\Delta X_{\rm B}^-)^2&=\tfrac12\left[\var{X_{\rm B}}+\var{X_{\rm I}'}-2\avg{X_{\rm B}X_{\rm I}'}\right]\nonumber\\
&=\tfrac12\left(\sigma_{\mathrm{en}}^{\rm B}+\sigma_{\rm J}^{\rm B}\right),
\end{align}
which reproduces Eq.~(\ref{eq:SNRbob}) with $\sigma_{\mathrm{en}}^{\rm B}=[\kappa_B+1-\eta_{\rm I}]N_{\mathrm{en}}$ and $\sigma_{\rm J}^{\rm B}$ given by Eqs.~(\ref{eq:jamclassic}) and (\ref{eq:jamquantum}); the ``$+1$'' in the latter is the sum of all vacuum contributions of the two measured modes. Eve's moments follow identically from Eq.~(\ref{eq:channel}) with $B\to E$, giving Eqs.~(\ref{eq:SNReve}) and (\ref{eq:jameve}); her vacuum budget is a single mode's worth, $1/2$.

Finally, describing Bob's and Eve's receptions by Eqs.~(\ref{eq:channel}) and (\ref{eq:noise}) with a shared noise mode $a_n$ is shorthand for a multiport network in which Eve's tap (transmissivity $\tau$) precedes the remaining propagation to Bob; the marginal moments of each receiver, which alone enter the SNRs, are identical, with $\eta_{\mathrm{ch},B}$ and $\eta_{\mathrm{ch},E}$ the end-to-end transmissivities of the respective paths. 

\section{Optimal linear combining}\label{app:gain}

Bob homodynes $X_{\rm B}$ and $X_{\rm I}'$ and forms $X_{\rm out}(g)=X_{\rm B}-gX_{\rm I}'$. Minimizing $V(g)=V_{\rm B}+g^2V_{\rm I}-2gC_{\rm BI}$ gives Eq.~(\ref{eq:gopt}). Because the estimator is formed electronically, no optical attenuation---and hence no added environmental noise---is required, and $g^\star$ may be tracked adaptively from the empirical covariance of the two photocurrents.

\textbf{Reduction to the conditional variance.} For $\eta_{\rm I}=1$, Eq.~(\ref{eq:VB}) collapses to $V_{\rm B}=\tfrac12+\eta_{\rm S}N_{\rm S}+\kappa_BN_{\mathrm{en}}$. With $V_{\rm I}=N_{\rm I}+\tfrac12=\var{X_{\rm I}}$ and $C_{\rm BI}=\sqrt{\eta_{\rm S}}\avg{X_{\rm S}X_{\rm I}}$, Eq.~(\ref{eq:gopt}) gives
\begin{equation}\label{eq:VminApp}
V_{\min}=\frac12+\kappa_BN_{\mathrm{en}}+\eta_{\rm S}N_{\rm S}-\frac{\eta_{\rm S}\avg{X_{\rm S}X_{\rm I}}^2}{N_{\rm I}+\tfrac12}.
\end{equation}
Subtracting the jamming-free reference $\tfrac12+\kappa_BN_{\mathrm{en}}$ and using $\var{X_{\rm S}}=N_{\rm S}+\tfrac12$,
\begin{align}
\Delta V&=\eta_{\rm S}\left[N_{\rm S}-\frac{\avg{X_{\rm S}X_{\rm I}}^2}{N_{\rm I}+\tfrac12}\right]\nonumber\\
&=\eta_{\rm S}\left[\var{X_{\rm S}}-\frac{\avg{X_{\rm S}X_{\rm I}}^2}{\var{X_{\rm I}}}-\frac12\right]\nonumber\\
&=\eta_{\rm S}\left[V_{{\rm S}|{\rm I}}-\frac12\right] ,
\end{align}
which is Eq.~(\ref{eq:master}) of the main text.

For a source with a nonnegative Glauber--Sudarshan $P$ function, the
quadratures can be decomposed as
\begin{equation}
X_{\rm S}=x_{\rm S}+\xi_{\rm S},
\qquad
X_{\rm I}=x_{\rm I}+\xi_{\rm I},
\end{equation}
where $x_{\rm S}$ and $x_{\rm I}$ are classical random amplitudes and
$\xi_{\rm S}$ and $\xi_{\rm I}$ are independent vacuum fluctuations, each
with variance $\tfrac{1}{2}$. Since $\xi_{\rm S}$ is uncorrelated with
$X_{\rm I}$,
\begin{equation}
\avg{X_{\rm S}X_{\rm I}}
=
\avg{x_{\rm S}x_{\rm I}},
\end{equation}
and the Cauchy--Schwarz inequality gives
\begin{equation}
\avg{x_{\rm S}x_{\rm I}}^2
\leq
\var{x_{\rm S}}\var{x_{\rm I}}
\leq
\var{x_{\rm S}}\var{X_{\rm I}} .
\end{equation}
Substituting into Eq.~(\ref{eq:condvardef}) yields
\begin{equation}
V_{{\rm S}|{\rm I}}^{\rm cl}
\geq
\var{x_{\rm S}}+\frac{1}{2}
-
\var{x_{\rm S}}
=
\frac{1}{2}.
\end{equation}

\textbf{The two sources.} Inserting $\avg{X_{\rm S}X_{\rm I}}^2=N_{\rm S}N_{\rm I}$ for the optimal classical source, and $N_{\rm S}(N_{\rm S}+1)$ with $N_{\rm I}=N_{\rm S}$ for the TMSV, gives Eqs.~(\ref{eq:residcl}) and (\ref{eq:residq}):
\begin{align}
\Delta V_{\rm cl}&=\eta_{\rm S}\left[N_{\rm S}-\frac{N_{\rm S}N_{\rm I}}{N_{\rm I}+\tfrac12}\right]
=\eta_{\rm S}N_{\rm S}\,\frac{1/2}{N_{\rm I}+\tfrac12}\nonumber\\
&=\frac{\eta_{\rm S}N_{\rm S}}{2N_{\rm I}+1},\\
\Delta V_{\rm q}&=\eta_{\rm S}\left[N_{\rm S}-\frac{N_{\rm S}(N_{\rm S}+1)}{N_{\rm S}+\tfrac12}\right]
=\eta_{\rm S}N_{\rm S}\,\frac{-1/2}{N_{\rm S}+\tfrac12}\nonumber\\
&=-\frac{\eta_{\rm S}N_{\rm S}}{2N_{\rm S}+1},
\end{align}
or equivalently
\begin{equation}\label{eq:condvars}
V_{{\rm S}|{\rm I}}^{\rm cl}=\frac12+\frac{N_{\rm S}}{2N_{\rm I}+1} ,
\qquad
V_{{\rm S}|{\rm I}}^{\rm q}=\frac{1}{4N_{\rm S}+2} .
\end{equation}
The two expressions differ only in the sign of the $\pm1/2$ generated in the
numerator---the vacuum unit that the classical correlation
$\sqrt{N_{\rm S}N_{\rm I}}$ falls short of and that the TMSV correlation
$\sqrt{N_{\rm S}(N_{\rm S}+1)}$ overshoots---but they are not mirror images
unless the classical idler is constrained to $N_{\rm I}=N_{\rm S}$. 

\textbf{Imperfect idler storage.} For general $\eta_{\rm I}$, $V_{\rm B}$ is
unchanged while $V_{\rm I}'=\eta_{\rm I}(N_{\rm I}+\tfrac12)+(1-\eta_{\rm
I})(N_{\mathrm{en}}+\tfrac12)$ and $C_{\rm BI}=\sqrt{\eta_{\rm S}\eta_{\rm
I}}\avg{X_{\rm S}X_{\rm I}}$, so Eq.~(\ref{eq:gopt}) gives
\begin{equation}\label{eq:DVgeneral}
\Delta V=\eta_{\rm S}\left[N_{\rm S}-\frac{\eta_{\rm I}\avg{X_{\rm S}X_{\rm I}}^2}{V_{\rm I}'}\right]
=\eta_{\rm S}\left[V_{{\rm S}|{\rm I}'}-\frac12\right] ,
\end{equation}
with $V_{{\rm S}|{\rm I}'}=\var{X_{\rm S}}-\avg{X_{\rm S}X_{\rm
I}'}^2/\var{X_{\rm I}'}$ the conditional variance given the stored
idler. Both bounds of Sec.~\ref{subsec:residual} therefore hold unchanged.
For the optimal classical source, $\avg{X_{\rm S}X_{\rm I}}^2=N_{\rm S}N_{\rm
I}$ and $\Delta V_{\rm cl}=\eta_{\rm S}N_{\rm S}[1-\eta_{\rm I}N_{\rm
I}/V_{\rm I}']\to0$ as $N_{\rm I}\to\infty$ at any fixed $\eta_{\rm I}>0$. For
the TMSV, $\avg{X_{\rm S}X_{\rm I}}^2=N_{\rm S}(N_{\rm S}+1)$ and $N_{\rm
I}=N_{\rm S}$, so $\Delta V_{\rm q}<0$ requires $\eta_{\rm I}N_{\rm S}(N_{\rm
S}+1)>N_{\rm S}V_{\rm I}'$; the common factor $N_{\rm S}$ cancels and the
condition collapses to
\begin{equation}
\tfrac12\eta_{\rm I}>(1-\eta_{\rm I})\big(N_{\mathrm{en}}+\tfrac12\big) ,
\end{equation}
i.e.\ Eq.~(\ref{eq:etaIthresh}), independently of $N_{\rm S}$. Taking $N_{\rm
S}\to\infty$ in Eq.~(\ref{eq:DVgeneral}) gives $\Delta V_{\rm q}\to\eta_{\rm
S}[(1-\eta_{\rm I})(N_{\mathrm{en}}+\tfrac12)/\eta_{\rm I}-\tfrac12]$, which is
Eq.~(\ref{eq:advloss}). At $\eta_{\rm I}=1$ these reduce to
Eqs.~(\ref{eq:residcl})--(\ref{eq:residq}).

\textbf{Optimal idler attenuation.} For general $\eta_{\rm I}$ and $N_{\mathrm{en}}=0$ the balanced receiver has $\sigma_{\rm J}^{\rm B}=\eta_{\rm S}N_{\rm S}+\eta_{\rm I}N_{\rm I}-2\sqrt{\eta_{\rm S}\eta_{\rm I}}\avg{X_{\rm S}X_{\rm I}}+1$. Setting $\partial\sigma_{\rm J}^{\rm B}/\partial\eta_{\rm I}=0$ gives $\sqrt{\eta_{\rm I}^\star}=\sqrt{\eta_{\rm S}}\avg{X_{\rm S}X_{\rm I}}/N_{\rm I}$, i.e.\ Eq.~(\ref{eq:etaIopt}), and back-substitution gives
\begin{equation}
\begin{aligned}
\sigma_{\rm J}^{\rm B}\big|_{\eta_{\rm I}^\star}
&=\eta_{\rm S}\left[N_{\rm S}-\frac{\avg{X_{\rm S}X_{\rm I}}^2}{N_{\rm I}}\right]+1\\[2pt]
&=\begin{cases}1 & \text{(classical)},\\[2pt] 1-\eta_{\rm S} & \text{(TMSV)},\end{cases}
\end{aligned}
\end{equation}
both exactly and for every $N_{\rm S}$, $N_{\rm I}$. Completing the square in
$\sqrt{\eta_{\rm I}}$ about this optimum gives Eq.~(\ref{eq:mismatch}) of the
main text. The classical result is independent of $N_{\rm I}$: an optical
attenuator admixes vacuum into the idler in exact proportion to the
attenuation, so this receiver cannot convert a brighter classical idler into a
better reference, and only the electronic combiner of Eq.~(\ref{eq:combiner})
can realize the optimized classical benchmark $\Delta V_{\rm cl}^{\rm opt}=0$.
Note also that $\eta_{\rm I}^\star=\eta_{\rm S}$ holds only for the symmetric
classical source; for the TMSV the optimum is $\eta_{\rm S}(N_{\rm
S}+1)/N_{\rm S}$, exceeding the matched value, and $\eta_{\rm I}=\eta_{\rm S}$
is optimal for the entangled source only as $N_{\rm S}\to\infty$. The balanced
matched scheme of the main text corresponds to $g^\star=1$; since
$V(g^\star)\le V(1)$ for every parameter choice, with the difference opening
only at second order in $g^\star-1$, the electronic combiner is never worse and
is insensitive to small errors in the estimated channel parameters.

\section{Wiretap-channel formulation}\label{app:wiretap}

This appendix places the SNR analysis of the main text in the framework of the classical wiretap channel \cite{Wyner1975,Csiszar1978,ElGamal2011} and shows that correlated jamming with a retained idler increases the secrecy capacity under an explicit, checkable condition.

\textbf{Setup.} Alice encodes message $M$ into codewords modulating the quadrature $X_{\rm A}$ over $n$ independent channel uses; Bob and Eve obtain the homodyne records described in the main text. Secrecy is measured by the normalized leakage $L(n)=\tfrac1n I(M;Z^n)$ and reliability by the error probability of Bob's decoder; a secrecy rate is achievable if both vanish as $n\to\infty$, and the supremum of achievable rates is the secrecy capacity \cite{Wyner1975,Csiszar1978}. For a degraded Gaussian wiretap channel with average power constraint, the secrecy capacity is the difference of the two channel capacities \cite{LeungYanCheong1978}, which for homodyne (real Gaussian) channels gives Eq.~(\ref{eq:secrecy_cap_SNR}).

\textbf{Degradedness of Eve's jammed channel.} Without jamming, Eve's record is $Z=\sqrt{\eta_{\mathrm{ch},E}}\,X_{\rm A}+\text{(Gaussian noise of variance }v)$; with jamming it is $Z'=\sqrt{\eta_{\mathrm{ch},E}}\,X_{\rm A}+\text{(Gaussian noise of variance }v'>v)$, with both noises independent of $X_{\rm A}$. Any additive Gaussian noise channel of larger noise variance is statistically degraded with respect to the smaller, in the sense that $Z'\stackrel{d}{=}Z+W$ with $W$ an independent Gaussian variable of variance $v'-v$. Hence $X_{\rm A}\to Z\to Z'$ is a Markov chain in distribution and the data-processing inequality gives
\begin{equation}\label{eq:DPI}
I(X_{\rm A};Z')\le I(X_{\rm A};Z) ,
\end{equation}
so that jamming can only reduce the information available to the eavesdropper.

\textbf{Bob's side and the net gain.} Bob's cancellation is not exact in general; define his self-jamming penalty and Eve's degradation,
\begin{equation}
\begin{aligned}
\Delta_B&=I(X_{\rm A};Y)-I(X_{\rm A};Y'),\\[2pt]
\Delta_E&=I(X_{\rm A};Z)-I(X_{\rm A};Z')\ge0,
\end{aligned}
\end{equation}
where $Y$ ($Y'$) is Bob's unjammed (jammed-and-cancelled) record and $\Delta_E\ge0$ follows from Eq.~(\ref{eq:DPI}). The quantity $\Delta_B$ carries no definite sign a priori: it is positive for the classical source and, as shown below, negative for the entangled one. The change in the achievable secrecy rate is
\begin{equation}\label{eq:netgain}
C'-C=\Delta_E-\Delta_B ,
\end{equation}
so jamming strictly helps whenever $\Delta_E>\Delta_B$. Equation~(\ref{eq:netgain}) is an identity for the unclipped rates $\tfrac12\log_2(1+\snr_{\rm B})-\tfrac12\log_2(1+\snr_{\rm E})$; wherever the clipping in Eq.~(\ref{eq:secrecy_cap_SNR}) is active its right-hand side is a lower bound on $C'-C$, which only strengthens the conclusions below.

The main text quantifies both contributions. $\Delta_E$ grows as
$\log(1+\eta_{\rm E}N_{\rm J})$, while $\Delta_B$ is controlled by
Eq.~(\ref{eq:master}) and carries the sign of $V_{{\rm S}|{\rm I}}-\tfrac12$.
For a classical source Eq.~(\ref{eq:clbound}) gives $\Delta_B>0$ strictly,
bounded above by the symmetric-benchmark value $\eta_{\rm S}N_{\rm
S}/(2N_{\rm S}+1)$ in the noise budget and tending to $0^+$ as the idler is
made bright; classical jamming is therefore asymptotically free for Bob, and
$C'>C$ follows from $\Delta_E>0$ alone. For the entangled source $\Delta
V_{\rm q}<0$ means Bob's post-cancellation noise drops below its unjammed
value, so $\Delta_B<0$ and jamming with a retained idler can only help Bob's
own channel, independently of Eve; in this limit $C'\ge C$ unconditionally.
This recovers the heuristic that Bob's idler acts as side information
unavailable to Eve, and shows that for the entangled source the jamming field
is not merely harmless to Bob but a resource he actively benefits from. We
emphasize that this statement refers to the optimal-combining receiver of
Sec.~\ref{subsec:combining}, for which Eq.~(\ref{eq:master}) applies; on the
balanced receiver used in Figs.~\ref{fig4} and \ref{fig5} the additional vacuum
port of the beam splitter keeps $\Delta_B>0$ for both sources. The two source
classes differ by the bounded margin $\mathcal{A}\le\eta_{\rm S}/2$ of
Eq.~(\ref{eq:advantage}), which is the entire information-theoretic content of
the entanglement in this protocol.

We would like to mention two caveats:  First, Bob's and Eve's noises are correlated through the shared jamming field, so the pair $(Y',Z')$ is not the product of independent channels. This is harmless. The secrecy capacity of a wiretap channel is a functional of the marginal conditional distributions $p(y|x)$ and $p(z|x)$ alone, and is invariant under any repackaging of the joint law $p(y,z|x)$ that preserves those marginals \cite{Csiszar1978,ElGamal2011}. The shorthand model of Eqs.~(\ref{eq:channel})--(\ref{eq:noise}) is therefore sufficient for the marginal rate calculation, although it is not a complete simultaneous bosonic dilation. What Eq.~(\ref{eq:secrecy_cap_SNR}) does assume is Gaussian codebooks and the stated homodyne receiver at Eve. Second, an unrestricted quantum Eve calls for replacing $I(X_{\rm A};Z')$ by the Holevo information of her conditional states. Since these are coherent states buried in thermal noise whose variance grows linearly in $N_{\rm J}$, the same mechanism suppresses her Holevo information; Appendix~\ref{app:holevo} makes this quantitative and shows that the substitution changes the numbers by well under one percent in the jammed regime of Figs.~\ref{fig4} and \ref{fig5}.

\section{Eve's Holevo information under a collective Gaussian attack}\label{app:holevo}

The main text grants Eve homodyne detection. Here we lift that restriction and
allow her the optimal collective measurement on her received modes, computing
her Holevo information $\chi_{\rm E}$ in closed form. Two facts make the
calculation tractable: the jamming field is refreshed symbol by symbol
(Sec.~\ref{subsec:setup}), so Eve's channel is memoryless and a single-symbol
quantity suffices; and every state involved is Gaussian.

\textbf{The ensemble.} Alice modulates the amplitude quadrature with a
zero-mean Gaussian of variance $\Sigma_{\rm A}=2N_{\rm A}$, sending the
coherent state $\vert\alpha\rangle$ with $\alpha=x/\sqrt2$ and $x\sim
\mathcal{N}(0,\Sigma_{\rm A})$, so that $\avg{|\alpha|^2}=N_{\rm A}$; this is
the modulation for which Eq.~(\ref{eq:SNReve}) is the Shannon SNR of the
induced real Gaussian channel. Propagating through
Eqs.~(\ref{eq:channel})--(\ref{eq:noise}), Eve's conditional state
$\rho_{{\rm E}|x}$ is a thermal state displaced by $\sqrt{\eta_{\mathrm{ch},E}}\,x$
along $X$, with quadrature variances
\begin{equation}\label{eq:Econd}
\begin{gathered}
\var{X_{\rm E}|x}=\var{P_{\rm E}|x}=\bar n_{\rm E}+\tfrac12 ,\\[2pt]
\bar n_{\rm E}=\eta_{\rm E}N_{\rm S}+\kappa_{\rm E}N_{\mathrm{en}} ,
\end{gathered}
\end{equation}
where $\eta_{\rm E}=\eta_n(1-\eta_{\mathrm{ch},E})$ and
$\kappa_{\rm E}=(1-\eta_{\mathrm{ch},E})(1-\eta_n)$, in agreement with
Eq.~(\ref{eq:jameve}). Both the jamming beam and the environment are phase
insensitive, so $\rho_{{\rm E}|x}$ carries no $X$--$P$ correlation and its
covariance matrix is $\bar n_{\rm E}+\tfrac12$ times the identity.

\textbf{Conditional entropy.} All conditional states are displaced copies of a
single thermal state of occupation $\bar n_{\rm E}$, and the von Neumann
entropy is displacement invariant, so
\begin{equation}\label{eq:Scond}
\begin{gathered}
\int\!dx\,p(x)\,S(\rho_{{\rm E}|x})=g(\bar n_{\rm E}) ,\\[2pt]
g(y)=(y{+}1)\log_2(y{+}1)-y\log_2 y ,
\end{gathered}
\end{equation}
with $g(0)=0$. The bosonic entropy function $g$ is strictly increasing and
strictly concave, with $g'(y)=\log_2\!\left(1+1/y\right)$; both properties are
used below. The value $g(0)=0$ is the vacuum limit.

\textbf{Average state and its symplectic eigenvalue.} Averaging over the
modulation adds $\eta_{\mathrm{ch},E}\Sigma_{\rm A}=2\eta_{\mathrm{ch},E}N_{\rm A}$
to the $X$ variance and nothing to $P$,
\begin{equation}\label{eq:Eavg}
\bar\sigma_{\rm E}=\mathrm{diag}\!\left(\bar n_{\rm E}+\tfrac12+2\eta_{\mathrm{ch},E}N_{\rm A},\;\;\bar n_{\rm E}+\tfrac12\right) ,
\end{equation}
a squeezed thermal state. Its symplectic eigenvalue is
$\nu_{\rm E}=\sqrt{\det\bar\sigma_{\rm E}}$, and using
$\snr_{\rm E}=2\eta_{\mathrm{ch},E}N_{\rm A}/(\bar n_{\rm E}+\tfrac12)$ from
Eq.~(\ref{eq:SNReve}) this collapses to the compact form
\begin{equation}\label{eq:nuE}
\nu_{\rm E}=\left(\bar n_{\rm E}+\tfrac12\right)\sqrt{1+\snr_{\rm E}} ,
\end{equation}
with $S(\bar\rho_{\rm E})=g(\nu_{\rm E}-\tfrac12)$. Hence
\begin{equation}\label{eq:chiE}
\boxed{\;\chi_{\rm E}=g\!\left(\nu_{\rm E}-\tfrac12\right)-g(\bar n_{\rm E})\;}
\end{equation}
with $\nu_{\rm E}$ given by Eq.~(\ref{eq:nuE}). Equation~(\ref{eq:chiE}) is
exact for all parameters. Setting $\bar n_{\rm E}\to\kappa_{\rm E}N_{\mathrm{en}}$
recovers the unjammed value.

Jamming suppresses $\chi_{\rm E}$ monotonically, and without calculation. Let
$\rho_{{\rm E}|x}^{(0)}$ denote Eve's conditional states in the absence of
jamming and $\rho_{{\rm E}|x}$ those with jamming. Since the jamming beam's
reduced state is thermal and independent of Alice's symbol, the two ensembles
are related by
$\rho_{{\rm E}|x}=\mathcal{N}_{\eta_{\rm E}N_{\rm S}}(\rho^{(0)}_{{\rm E}|x})$,
where $\mathcal{N}_{\bar n}$ is the additive classical-noise channel applying a
random Gaussian displacement of variance $\bar n$ per quadrature.
$\mathcal{N}_{\bar n}$ is a fixed CPTP map, applied identically to every member
of the ensemble and independent of $x$, so the data-processing inequality for
the Holevo quantity gives
\begin{equation}\label{eq:chiDPI}
\chi_{\rm E}\big(\{p(x),\mathcal{N}_{\bar n}(\rho^{(0)}_{{\rm E}|x})\}\big)\le\chi_{\rm E}\big(\{p(x),\rho^{(0)}_{{\rm E}|x}\}\big) ,
\end{equation}
and, by composing $\mathcal{N}_{\bar n}\circ\mathcal{N}_{\bar n'}=\mathcal{N}_{\bar n+\bar n'}$,
$\chi_{\rm E}$ is nonincreasing in $\bar n_{\rm E}$ and hence in the jamming
brightness $N_{\rm S}$. This is the quantum counterpart of
Eq.~(\ref{eq:DPI}) and requires no evaluation of Eq.~(\ref{eq:chiE}).

\textbf{Two-sided bounds.} Write $\nu_{\rm E}-\tfrac12=\bar n_{\rm E}+\delta$
with
\begin{equation}\label{eq:delta}
\delta=\left(\bar n_{\rm E}+\tfrac12\right)\left(\sqrt{1+\snr_{\rm E}}-1\right)\;>0 .
\end{equation}
Concavity of $g$ then brackets Eq.~(\ref{eq:chiE}) rigorously,
\begin{equation}\label{eq:chibounds}
\delta\,\log_2\!\left(1+\frac{1}{\bar n_{\rm E}+\delta}\right)\;\le\;\chi_{\rm E}\;\le\;\delta\,\log_2\!\left(1+\frac{1}{\bar n_{\rm E}}\right) .
\end{equation}
Both bounds vanish as $\bar n_{\rm E}\to\infty$ at fixed $\delta$; the upper
one already establishes $\chi_{\rm E}\to0$ under strong jamming, at the rate
$\chi_{\rm E}\simeq\eta_{\mathrm{ch},E}N_{\rm A}/(\bar n_{\rm E}\ln2)$, i.e.\
inversely in the jamming brightness.

\begin{figure}[t!]
\centering{\includegraphics[width=1\columnwidth]{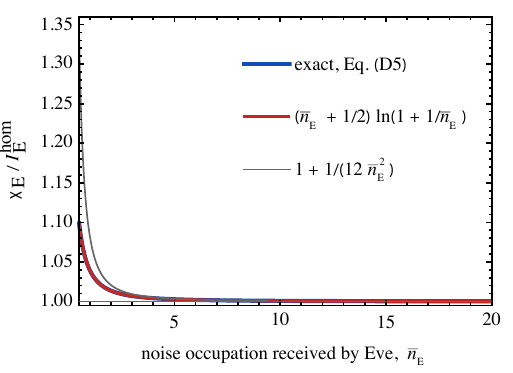}}
\caption{Ratio of Eve's Holevo information to her homodyne information in the
strongly jammed limit, from Eq.~(\ref{eq:chiE}), against the noise occupation
$\bar n_{\rm E}$ she receives, together with the asymptotic form of
Eq.~(\ref{eq:chiratio}) and its leading correction $1+1/(12\bar n_{\rm
E}^{2})$. The advantage of a collective measurement over simple homodyne
detection diverges for an unjammed coherent-state ensemble as $\bar n_{\rm
E}\to0$, and is closed by jamming as the inverse square of the noise Eve
receives. The exact curve is evaluated at $\snr_{\rm E}=10^{-4}$, deep in the
strongly jammed limit, for $\bar n_{\rm E}$ from $0.5$ to $20$; at $\bar n_{\rm
E}=5$ the exact ratio is $1.0028$ against $1.0033$ from the leading correction,
and at $\bar n_{\rm E}=20$ it is $1.00020$ against $1.00021$. This is the
quantitative content of assumption (iii) of Sec.~\ref{subsec:threat}.}
\label{fig6}
\end{figure}

\textbf{Strong-jamming asymptotics: the collective attack.} The
regime of interest is $\snr_{\rm E}\ll1$, which is what jamming is for.
Expanding Eq.~(\ref{eq:delta}), $\delta=\tfrac12(\bar n_{\rm E}+\tfrac12)\snr_{\rm E}+O(\snr_{\rm E}^2)
=\eta_{\mathrm{ch},E}N_{\rm A}+O(\snr_{\rm E}^2)$, so
$\chi_{\rm E}\to\delta\,g'(\bar n_{\rm E})$, while the homodyne value is
$I_{\rm E}^{\rm hom}=\tfrac12\log_2(1+\snr_{\rm E})\to\snr_{\rm E}/(2\ln2)$.
The ratio is independent of $N_{\rm A}$ and of $\eta_{\mathrm{ch},E}$,
\begin{equation}\label{eq:chiratio}
\begin{aligned}
\frac{\chi_{\rm E}}{I_{\rm E}^{\rm hom}}\;&\longrightarrow\;
\left(\bar n_{\rm E}+\tfrac12\right)\ln\!\left(1+\frac{1}{\bar n_{\rm E}}\right)\\[2pt]
&=1+\frac{1}{12\,\bar n_{\rm E}^{2}}-\frac{1}{12\,\bar n_{\rm E}^{3}}+O(\bar n_{\rm E}^{-4}) .
\end{aligned}
\end{equation}
This is the central result of this appendix, and it is plotted in
Fig.~\ref{fig6}. The gap between a collective attack and simple homodyne
detection---unbounded for an unjammed coherent-state ensemble, where
$\bar n_{\rm E}\to0$ and the left-hand side of Eq.~(\ref{eq:chiratio})
diverges logarithmically---is closed by jamming as the inverse square of the
noise Eve receives. Physically, the jamming thermalizes Eve's conditional
states until they are so classical that no collective measurement can extract
appreciably more than a quadrature measurement. The homodyne approximation is therefore quantitatively accurate in the
strong-jamming regime, although the collective-measurement bound remains the
appropriate conservative description.

\textbf{Achievable secrecy rate.} Against collective attacks the
Devetak--Winter rate $I(X_{\rm A};Y)-\chi_{\rm E}$ is achievable
\cite{DevetakWinter2005}, so with Bob's actual (homodyne) receiver
\begin{equation}\label{eq:CHol}
C^{\rm Hol}=\max\!\left[0,\;\tfrac12\log_2\!\left(1+\snr_{\rm B}\right)-\chi_{\rm E}\right] ,
\end{equation}
which lower-bounds the secrecy capacity, since Bob is not optimized over
measurements. Comparing Eqs.~(\ref{eq:CHol}) and (\ref{eq:secrecy_cap_SNR}),
the entire cost of upgrading Eve from homodyne to collective is the excess of
$\chi_{\rm E}$ over $\tfrac12\log_2(1+\snr_{\rm E})$, bounded by
Eq.~(\ref{eq:chiratio}).

\textbf{Source independence.} Because Eve's reduced state is thermal for both
sources (Sec.~\ref{subsec:sources}), Eqs.~(\ref{eq:Econd})--(\ref{eq:CHol})
depend on the noise source only through $N_{\rm S}$. The quantum advantage
$\mathcal{A}$ of Eq.~(\ref{eq:advantage}) therefore carries over to the
collective-attack setting unchanged: it lives entirely in $\snr_{\rm B}$, and
$\chi_{\rm E}$ is common to both strategies. If anything it is more visible
here, because it is largest in relative terms near the secrecy cutoff, where
$C^{\rm Hol}$ is small.

\textbf{Numerical evaluation.} Table~\ref{tab:holevo} evaluates
Eqs.~(\ref{eq:chiE}) and (\ref{eq:CHol}) at the parameters of
Fig.~\ref{fig4}, with Bob using the matched balanced receiver and a symmetric
classical source. Across the whole secure region the collective attack costs
less than $0.5\%$ of Eve's information and shifts the secrecy cutoff from
$\eta_{\mathrm{ch},E}=0.712$ to $0.710$. Without jamming the same upgrade
costs $12\%$--$16\%$ and eliminates the residual secrecy outright: at
$\eta_{\mathrm{ch},E}=0.299$ the unjammed capacity falls from
$2.4\times10^{-3}$ to zero, while the jammed capacity falls only from
$0.9637$ to $0.9625$. Correlated jamming thus does something the SNR analysis
alone does not reveal---it renders the achievable secrecy rate nearly independent of
which adversary model is assumed. Replacing the classical source by the TMSV
raises $C^{\rm Hol}$ from $0.9625$ to $1.0980$ at
$\eta_{\mathrm{ch},E}=0.299$ and from $0.0267$ to $0.1621$ at
$\eta_{\mathrm{ch},E}=0.7$.

\begin{table*}[t]
\caption{Eve's information under homodyne detection and under an optimal
collective measurement, from Eqs.~(\ref{eq:SNReve}) and (\ref{eq:chiE}), and
the resulting secrecy rates from Eqs.~(\ref{eq:secrecy_cap_SNR}) and
(\ref{eq:CHol}). Parameters as in Fig.~\ref{fig4}: $N_{\rm A}=20$,
$N_{\mathrm{en}}=0.1$, $\eta_{\mathrm{ch},B}=0.3$, $\eta_n=0.3$, with Bob on
the matched balanced receiver. Upper block: jamming with a symmetric source of
brightness $N_{\rm J}=25$ ($\tfrac12\log_2(1+\snr_{\rm B})=1.7704$~bit).
Lower block: no jamming ($1.7704\to2.2573$~bit). Rates in bits per channel
use.}
\label{tab:holevo}
\begin{ruledtabular}
\begin{tabular}{cccccccc}
$\eta_{\mathrm{ch},E}$ & $\bar n_{\rm E}$ & $\snr_{\rm E}$ & $I_{\rm E}^{\rm hom}$ & $\chi_{\rm E}$ & $\chi_{\rm E}/I_{\rm E}^{\rm hom}$ & $C$ [Eq.~(\ref{eq:secrecy_cap_SNR})] & $C^{\rm Hol}$ [Eq.~(\ref{eq:CHol})] \\
\colrule
\multicolumn{8}{c}{\textbf{with correlated jamming, $N_{\rm J}=25$}}\\
0.100 & 6.813 & 0.547 & 0.3147 & 0.3151 & 1.0013 & 1.4557 & 1.4553 \\
0.200 & 6.056 & 1.220 & 0.5754 & 0.5761 & 1.0013 & 1.1950 & 1.1943 \\
0.299 & 5.307 & 2.060 & 0.8067 & 0.8079 & 1.0015 & 0.9637 & 0.9625 \\
0.400 & 4.542 & 3.173 & 1.0306 & 1.0324 & 1.0018 & 0.7398 & 0.7380 \\
0.500 & 3.785 & 4.667 & 1.2513 & 1.2541 & 1.0022 & 0.5191 & 0.5163 \\
0.600 & 3.028 & 6.803 & 1.4820 & 1.4862 & 1.0029 & 0.2884 & 0.2842 \\
0.700 & 2.271 & 10.105 & 1.7365 & 1.7437 & 1.0041 & 0.0339 & 0.0267 \\
\colrule
\multicolumn{8}{c}{\textbf{without jamming}}\\
0.100 & 0.0630 & 7.11 & 1.5094 & 1.7542 & 1.162 & 0.7479 & 0.5031 \\
0.200 & 0.0560 & 14.39 & 1.9719 & 2.2390 & 1.135 & 0.2854 & 0.0183 \\
0.299 & 0.0491 & 21.78 & 2.2549 & 2.5379 & 1.126 & 0.0024 & 0.0000 \\
\end{tabular}
\end{ruledtabular}
\end{table*}

Equation~(\ref{eq:CHol}) grants Eve the optimal
joint measurement on the modes she collects, but retains assumptions (i),
(ii), and (iv) of Sec.~\ref{subsec:threat}: her collection remains bounded,
she injects nothing, and she never obtains the idler. In particular she is not
given the purification of the channel, which is what separates this analysis
from a QKD security proof; and the treatment is a collective-attack, not a
coherent-attack, bound, the latter requiring de Finetti or entropic-uncertainty
machinery \cite{Weedbrook2012,Pirandola2020}. Within the physical-layer threat
model of Sec.~\ref{sec:security}, however, Eq.~(\ref{eq:chiratio}) shows that
the choice of measurement model is no longer where the uncertainty lies.

\end{document}